\documentclass[twocolumn,10pt]{asme2ej}
\usepackage{amssymb}
\usepackage{amsfonts}
\usepackage{amsmath}
\usepackage{euscript}
\usepackage{color}
\usepackage{graphicx}
\usepackage{rotating}
\usepackage{bm}
\usepackage{curves}
\usepackage[svgnames,dvipsnames]{pstricks}
\usepackage{lettrine}
\usepackage{mathrsfs}
\usepackage{booktabs}
\usepackage{dcolumn}
\DeclareGraphicsExtensions{{},.eps}
\providecommand{\tsc}[1]{{\text{\sc#1}}}

\providecommand{\etal}{et al.}
\providecommand{\tabref}[1]{{\textup{(Tab.~\ref{#1})}}}
\providecommand{\figref}[1]{{\textup{(Fig.~\ref{#1})}}}

\providecommand{\figrefnp}[1]{{\textup{Fig.~\ref{#1}}}}

\providecommand{\eqrefsatob} [2]{\textup{(\ref{#1}--\ref{#2})}}
\providecommand{\eqrefsab}   [2]{\textup{(\ref{#1},\ref{#2})}}
\providecommand{\eqrefsabc}  [3]{\textup{(\ref{#1},\ref{#2},\ref{#3})}}
\providecommand{\eqrefsabcd} [4]{\textup{(\ref{#1},\ref{#2},\ref{#3},\ref{#4})}}

\providecommand{\tabrefsab}   [2]{{\textup{(Tabs.~\ref{#1},\ref{#2})}}}
\providecommand{\tabrefsatob} [2]{{\textup{(Tabs.~\ref{#1}--\ref{#2})}}}

\newcommand{\abs}[1]{\lvert#1\rvert}

\newcommand{\tsr}[1]{{\pmb{\mathsf{#1}}}}

\newcommand{\tr}{\mathrm{tr}}
   \newcommand{\I}[1]{  \mathrm{I}_{\pmb{\mathsf{#1}}}}
  \newcommand{\II}[1]{ \mathrm{II}_{\pmb{\mathsf{#1}}}}
 \newcommand{\III}[1]{\mathrm{III}_{\pmb{\mathsf{#1}}}}
   \newcommand  {\Ia}[3]{  \mathrm{I}_{\pmb{\mathsf{#1}}^{#2}_{#3}}}
   \newcommand {\Iab}[6]{  \mathrm{I}_{\pmb{\mathsf{#1}}^{#2}_{#3}\pmb{\mathsf{#4}}^{#5}_{#6}}}
   
\newcommand{\devnd}[1]{\overset{_\star}{\pmb{\mathsf{#1}}}}
\newcommand{\tsrbar}[1]{\bar{\pmb{\mathsf{#1}}}}
\newcommand{\strd}[1]{\overset{_\star}{#1}}
\title{Tensorial representations of Reynolds-stress pressure-strain redistribution}
\author{G.A. Gerolymos, C. Lo, I. Vallet
        \affiliation{Universit\'e Pierre-et-Marie-Curie {\textup(\tsc{upmc})}, 4 place Jussieu, 75005 Paris, France\\
                     Emails: georges.gerolymos@upmc.fr, celine.lo@upmc.fr, isabelle.vallet@upmc.fr\\
                     \date{\today}
                    }
       }
\begin{document}
\maketitle
\begin{abstract}
{\em
The purpose of the present note is to contribute in clarifying the relation between representation bases used in the closure for the redistribution (pressure-strain) tensor $\phi_{ij}$,
and to construct representation bases whose elements have clear physical significance.
The representation of different models in the same basis is essential for comparison purposes,
and the definition of the basis by physically meaningfull tensors adds insight to our understanding of closures.
The rate-of-production tensor can be split into production by mean strain and production by mean rotation $P_{ij}=P_{\bar S_{ij}}+P_{\bar\Omega_{ij}}$.
The classic representation basis $\mathfrak{B}[\tsr{b}, \tsrbar{S}, \tsrbar{\Omega}]$ of homogeneous turbulence
{\em [{\em eg} Ristorcelli J.R., Lumley J.L., Abid R.: {\it J. Fluid Mech.} {\bf 292} (1995) 111--152]},
constructed from the anisotropy $\tsr{b}$, the mean strain-rate $\tsrbar{S}$, and the mean rotation-rate $\tsrbar{\Omega}$ tensors,
is interpreted, in the present work, in terms of the relative contributions of the deviatoric tensors
$P^{(\mathrm{dev})}_{\bar S_{ij}}:=P_{\bar S_{ij}}-\tfrac{2}{3}P_\mathrm{k}\delta_{ij}$ and $P^{(\mathrm{dev})}_{\bar\Omega_{ij}}:=P_{\bar\Omega_{ij}}$.
Different alternative equivalent representation bases, explicitly using $P^{(\mathrm{dev})}_{\bar S_{ij}}$ and $P_{\bar\Omega_{ij}}$ are discussed,
and the projection rules between bases are caclulated, using a matrix-based systematic procedure.
An initial term-by-term {\em a priori} investigation of different second-moment closures is undertaken.
}
\end{abstract}
%
%
%
%
%
%
\section{Introduction}\label{TRRSPSR_s_I}
%
%
%
%
%

The pressure-strain (redistribution) correlation \cite{Rotta_1951a,
                                                       Reynolds_1974a,
                                                       Lumley_1978a}
\begin{alignat}{6}
\phi_{ij}:=2\overline{p'S_{ij}'}=\phi^{({\rm r};{\mathfrak V})}_{ij}+\phi^{({\rm s};{\mathfrak V})}_{ij}+\phi^{(w)}_{ij}
                                                                                                                                    \label{Eq_TRRSPSR_001}
\end{alignat}
"{\em plays a pivotal role in determining the structure of a wide variety of turbulent flows}" \cite{Speziale_Sarkar_Gatski_1991a}.
In \eqref{Eq_TRRSPSR_001}, $p$ is the pressure, $S_{ij}:=\tfrac{1}{2}(\partial_{x_j} u_i+\partial_{x_i} u_j)$ is the rate-of-strain tensor,
$u_i$ are the velocity-components in the reference-frame with coordinates $x_i$,
$\bar{(\cdot)}$ denotes ensemble averaging, $(\cdot)'$ denotes turbulent fluctuations,
$\phi^{({\rm r};{\mathfrak V})}_{ij}$ is the rapid (meanflow/turbulence interaction) volume-part of redistribution \cite{Speziale_Sarkar_Gatski_1991a},
$\phi^{({\rm s};{\mathfrak V})}_{ij}$ is the slow (turbulence/turbulence interaction) volume-part of redistribution \cite{Speziale_Sarkar_Gatski_1991a},
and $\phi^{(w)}_{ij}:=\phi^{({\rm r};w)}_{ij}+\phi^{({\rm s};w)}_{ij}$ are the wall-echo terms near solid boundaries \cite{Chou_1945a},
obtained by the free-space Green's function solution of the incompressible flow
Poisson equation (linear in $p'$) for the fluctuating pressure \cite{Chou_1945a}.
The most general approach to modelling $\phi_{ij}$ in homogeneous turbulence is based
on the order-4 tensors associated with the volume integrals $\phi^{({\rm r};{\mathfrak V})}_{ij}$ and $\phi^{({\rm s};{\mathfrak V})}_{ij}$ \cite{Ristorcelli_Lumley_Abid_1995a},
which require in particular, through obvious scaling arguments,
that retained models for $\phi_{ij}^{(r)}$ should be linear in meanflow-velocity gradients $\partial_{x_j}\bar u_i$ \cite{Lumley_1978a},
and that $\phi_{ij}^{(s)}$ should not depend directly on $\partial_{x_j}\bar u_i$ \cite{Lumley_1978a}.
The redistribution tensor is symmetric by definition \eqref{Eq_TRRSPSR_001} and deviatoric,
because of the incompressible fluctuating continuity equation
($\phi_{\ell\ell}\stackrel{\eqref{Eq_TRRSPSR_001}}{=}2\overline{p'\partial_{x_\ell} u'_\ell}=0$).
Therefore second-moment closures (\tsc{smc}s) represent redistribution as a linear combination of deviatoric symmetric tensors \cite{Speziale_Sarkar_Gatski_1991a,
                                                                                                                                     Ristorcelli_Lumley_Abid_1995a,
                                                                                                                                     Fu_Wang_1997a}.
Let
\begin{subequations}
                                                                                                                                    \label{Eq_TRRSPSR_002}
\begin{alignat}{6}
\bar S_{ij}          :=&\tfrac{1}{2}\left(\partial_{ x_j}\bar u_i+\partial_{x_i}\bar u_j\right)
                                                                                                                                    \label{Eq_TRRSPSR_002a}\\
\bar\Omega_{ij}      :=&\tfrac{1}{2}\left(\partial_{ x_j}\bar u_i-\partial_{x_i}\bar u_j\right) \stackrel{\eqref{Eq_TRRSPSR_002a}}{=}\partial_{ x_j}\bar u_i-\bar S_{ij}
                                                                                                                                    \label{Eq_TRRSPSR_002b}\\
b_{ij} :=&\big(\overline{u'_\ell u'_\ell}\big)^{-1}\;\overline{u'_i u'_j}-\tfrac{1}{3}\delta_{ij}
                                                                                                                                    \label{Eq_TRRSPSR_002c}\\
P_{\bar S_{ij}}      :=&-\rho\overline{u'_i u'_\ell}\bar      S_{j\ell}-\rho\overline{u'_j u'_\ell}\bar      S_{i\ell}
                                                                                                                                    \label{Eq_TRRSPSR_002d}\\
P_{\bar \Omega_{ij}} :=&-\rho\overline{u'_i u'_\ell}\bar \Omega_{j\ell}-\rho\overline{u'_j u'_\ell}\bar \Omega_{i\ell}
                                                                                                                                    \label{Eq_TRRSPSR_002e}\\
P_{ij}                :=&-\rho\overline{u'_i u'_\ell}\partial_{x_\ell}\bar u_j-\rho\overline{u'_j u'_\ell}\partial_{x_\ell}\bar u_i\stackrel{\eqrefsatob{Eq_TRRSPSR_002a}
                                                                                                                                                        {Eq_TRRSPSR_002e}}{=}P_{\bar S_{ij}}+P_{\bar\Omega_{ij}}
                                                                                                                                    \label{Eq_TRRSPSR_002f}\\
D_{ij}                :=&-\rho\overline{u'_i u'_\ell}\partial_{x_j}\bar u_\ell-\rho\overline{u'_j u'_\ell}\partial_{x_i}\bar u_\ell\stackrel{\eqrefsatob{Eq_TRRSPSR_002a}
                                                                                                                                                        {Eq_TRRSPSR_002e}}{=}P_{\bar S_{ij}}-P_{\bar\Omega_{ij}}
                                                                                                                                    \label{Eq_TRRSPSR_002g}
\end{alignat}
\end{subequations}
{\em viz}
the Reynolds-stress-anisotropy tensor $b_{ij}$ \eqref{Eq_TRRSPSR_002c} \cite{Lumley_1978a},
the production tensor $P_{ij}$ \eqref{Eq_TRRSPSR_002f} appearing in Reynolds-stress transport \cite{Speziale_Sarkar_Gatski_1991a},
which can be separated \eqref{Eq_TRRSPSR_002f} in production by mean strain-rate $P_{\bar S_{ij}}$ \eqref{Eq_TRRSPSR_002d}
and production by mean rotation-rate $P_{\bar\Omega_{ij}}$ \eqref{Eq_TRRSPSR_002e}.
When homogeneous turbulence is submitted to irrotational strain $P_{\bar\Omega_{ij}}=0$, while in case of solid-body rotation $P_{\bar S_{ij}}=0$.
The tensor $D_{ij}$ \eqref{Eq_TRRSPSR_002g}, which bears no particular name, appears in several early pressure-strain closures \cite{Reynolds_1974a,
                                                                                                                                   Naot_Shavit_Wolfshtein_1973a,
                                                                                                                                   Launder_Reece_Rodi_1975a}.
Naot \etal~\cite{Naot_Shavit_Wolfshtein_1973a} attribute to Reynolds \cite{Reynolds_1974a} its introduction in models for $\phi_{ij}$.
By \eqref{Eq_TRRSPSR_002g}, it follows that $D_{ij}$ can also be interpreted, {\em indeed should be defined}, as the difference of production by stain minus production by rotation.

Almost all practical models\footnote{\label{ff_TRRSPSR_001}Notice that Launder \cite{Launder_1989a} has suggested to further include the advection tensor $C_{ij}:=\rho D_t\overline{u_i'u_j'}$ in the representation,
                                                           but this has not become standard practice. 
                                                           In the case of spatially evolving stationary quasi-homogeneous turbulence, $C_{ij}=\rho\bar u_\ell\partial_{x_\ell}\overline{u_i'u_j'}\neq0$,
                                                           while in \tsc{dns} studies of time-evolving spatially homogeneous turbulence, $C_{ij}=\rho\partial_t\overline{u_i'u_j'}\neq0$,
                                                           implying that the suggestion of including $C_{ij}$ in the representation merits further study.
                                    }
for $\phi_{ij}$ \cite{Rotta_1951a,
                      Reynolds_1974a,
                      Lumley_1978a,
                      Speziale_Sarkar_Gatski_1991a,
                      Ristorcelli_Lumley_Abid_1995a,
                      Fu_Wang_1997a,
                      Naot_Shavit_Wolfshtein_1973a,
                      Launder_Reece_Rodi_1975a,
                      Naot_Shavit_Wolfshtein_1970a,
                      Dafalias_Younis_2009a}
correspond to a linear combination of tensors in \eqref{Eq_TRRSPSR_002} and/or tensors constructed by products of tensors in \eqref{Eq_TRRSPSR_002} which are linear in the mean-velocity gradients.
The particular models range from simple phenomenological representations \cite{Rotta_1951a,
                                                                               Naot_Shavit_Wolfshtein_1970a,
                                                                               Naot_Shavit_Wolfshtein_1973a,
                                                                               Reynolds_1974a,
                                                                               Launder_Reece_Rodi_1975a}
using only tensors in \eqref{Eq_TRRSPSR_002},
to more complex models making reference to irreducible representation bases \cite{Speziale_Sarkar_Gatski_1991a,
                                                                                  Ristorcelli_Lumley_Abid_1995a,
                                                                                  Fu_Wang_1997a}.
It is well established that the most general models can be represented in a basis of 8 linearly independent tensors \cite{Ristorcelli_Lumley_Abid_1995a}.
However, there are several possible choices of the basis-elements, and the originally established forms of different models \cite{Rotta_1951a,
                                                                                                                                  Reynolds_1974a,
                                                                                                                                  Lumley_1978a,
                                                                                                                                  Speziale_Sarkar_Gatski_1991a,
                                                                                                                                  Ristorcelli_Lumley_Abid_1995a,
                                                                                                                                  Fu_Wang_1997a,
                                                                                                                                  Naot_Shavit_Wolfshtein_1973a,
                                                                                                                                  Launder_Reece_Rodi_1975a,
                                                                                                                                  Naot_Shavit_Wolfshtein_1970a,
                                                                                                                                  Dafalias_Younis_2009a}
use different basis-elements. Although most of the models have been projected on a common basis \cite{Ristorcelli_Lumley_Abid_1995a},
the most complex one \cite{Fu_1988a,
                           Craft_Launder_2001a}
has invariably been expressed in a reducible form,
including tensors which are linearly dependent \cite{Ristorcelli_Lumley_Abid_1995a}
and can be projected on an 8-element basis \cite{Lo_2011a}.
To make detailed comparisons between different models, going beyond global evaluation of results against data for a given testcase,
it is necessary to express the models in a common basis. The purpose of the present work is to contribute in advancing towards
the answer to the questions:
1) what is the common representation basis that should be retained,
2) what is the physical significance of the basis-elements, and
3) how different modelling choices associated with different routes followed in the construction of various models can be compared.

In \S\ref{TRRSPSR_s_CPRB} we summarize results concerning the classical 8-element representation basis \cite{Ristorcelli_Lumley_Abid_1995a},
and debate on the arguments in favour of a polynomial representation basis {\em vs} a functional representation basis \cite{Speziale_Sarkar_Gatski_1991a,
                                                                                                                           Ristorcelli_Lumley_Abid_1995a}.
In \S\ref{TRRSPSR_s_PSR} we reinterpret the classical representation basis in terms of production by mean strain-rate $P_{\bar S_{ij}}$ \eqref{Eq_TRRSPSR_002d}
and production by mean rotation-rate $P_{\bar\Omega_{ij}}$ \eqref{Eq_TRRSPSR_002e}, illustrating the physical significance to the 3 basis elements, which represent all
quasilinear models for rapid pressure-strain \cite{Shima_1998a}.
In \S\ref{TRRSPSR_s_SRB} we consider alternative bases built from the symmetric tensors $[b_{ij},P_{ij},D_{ij}]$ or $[b_{ij},P_{\bar S_{ij}}, P_{\bar \Omega_{ij}}]$,
which under the requirement of linearity in mean-velocity gradients are symmetric in $[P_{ij},D_{ij}]$ or $[P_{\bar S_{ij}}, P_{\bar \Omega_{ij}}]$, respectively,
we show that the mean rate-of-strain tensor $\bar S_{ij}$ can be explicitly projected on these bases, in which it does not appear explicitly,
and discuss the advantages and drawbacks of such symmetric bases generated from tensors appearing in the transport equations for the Reynolds-stresses.
In \S\ref{TRRSPSR_s_CoBR} we briefly note how projection of models on different bases can be made systematic using projection matrices and their inverses.
Finally in \S\ref{TRRSPSR_s_D}, we illustrate, through {\em a priori} analysis of \tsc{dns} data for fully developed plane channel flow, how term-by-term comparison of various
models, expressed in a common basis, can be used to highlight different modelling strategies, indicating directions for future work (\S\ref{TRRSPSR_s_E}).
                      
%
%
%
%
%
\section{Classical polynomial representation basis}\label{TRRSPSR_s_CPRB}
%
%
%
%
%

Recall \cite{Rivlin_1955a},
that every order-2 tensor $\tsr{A}\in\mathbb{E}^{3\times3}$ has 3 invariants,
$\I{A}:=\tr\tsr{A}$, $\II{A}:=\tfrac{1}{2}(\tr^2\tsr{A}-\tr\tsr{A}^2)=\tfrac{1}{2}\left((A_{\ell\ell})^2-A_{\ell m}A_{\ell m}\right)$,
and $\III{A}:=\det\tsr{A}=\epsilon_{ijk}A_{1i}A_{2j}A_{3k}=\epsilon_{ijk}A_{i1}A_{j2}A_{k3}$,
coefficients of its characteristic polynomial $\mathbb{R}[x]\ni p(x;\tsr{A}):=x^3-\I{A} x^2 +\II{A} x -\III{A}$,
whose roots $\in\mathbb{C}$ (or $\in\mathbb{R}$ if $\tsr{A}$ is symmetric) are the eigenvalues of $\tsr{A}$,
and which, by the Cayley-Hamilton theorem, is satisfied by $\tsr{A}$, {\em ie}
$\tsr{A}^3= \I{A} \tsr{A}^2 -\II{A} \tsr{A} +\III{A} \tsr{I}_3$ \cite[(4.24), p. 689]{Rivlin_1955a}.
Applying the Cayley-Hamilton theorem to $(\tsr{A}+\tsr{B})^3-(\tsr{A}-\tsr{B})^3$ yields the Cayley-Hamilton theorem extension
$\tsr{ABA}=-\tsr{A}^2\tsr{B}
-\tsr{B}^2\tsr{A}
+\I{B}\tsr{A}^2
+(\I{AB}-\I{A}\I{B})\tsr{A}
-\II{A}\tsr{B}
+\I{A}\tsr{AB}
+\I{A}\tsr{BA}
+(\Iab{A}{2}{}{B}{}{}-\I{A}\I{AB}+\I{B}\II{A})\tsr{I}_3$
\cite[(4.22), p. 688]{Rivlin_1955a}, and applying it to $(\tsr{A}+\tsr{B}+\tsr{C})^3$
results in the extended Cayley-Hamilton theorem
$\tsr{ABC}+\tsr{CBA}+\tsr{BCA}+\tsr{ACB}+\tsr{CAB}+\tsr{BAC}= (\I{BC}-\I{B}\I{C})\tsr{A}
                                                             +(\I{CA}-\I{C}\I{A})\tsr{B}
                                                             +(\I{AB}-\I{A}\I{B})\tsr{C}
                                                             +\I{A}\tsr{BC}
                                                             +\I{A}\tsr{CB}
                                                             +\I{B}\tsr{CA}
                                                             +\I{B}\tsr{AC}
                                                             +\I{C}\tsr{BA}
                                                             +\I{C}\tsr{AB}
                                                             +(\I{ABC}+\I{CBA}-\I{A}\I{BC}-\I{B}\I{CA}+\I{C}\I{AB}+\I{A}\I{B}\I{C})\tsr{I}_3$
\cite[(4.21), p. 688]{Rivlin_1955a}.
The operator $\I{(.)}:=\mathrm{tr}(\cdot)$ denotes the trace (first invariant) of a tensor, and
$\tsr{I}_3\in\mathbb{E}^{3\times3}$ is the order-2 identity tensor in the Euclidean space $\mathbb{E}^3$.
With respect to $\tsr{b}$ \eqref{Eq_TRRSPSR_002c}, $\tsrbar{S}$ \eqref{Eq_TRRSPSR_002a}
and $\tsrbar{\Omega}$ \eqref{Eq_TRRSPSR_002b}, the Cayley-Hamilton theorem and its extensions \cite{Rivlin_1955a}
give \cite{Speziale_Sarkar_Gatski_1991a,
           Ristorcelli_Lumley_Abid_1995a}
\begin{subequations}
                                                                                                                                    \label{Eq_TRRSPSR_003}
\begin{alignat}{6}
\tsr{b}^3=&-\II{b}\tsr{b}+\III{b}\tsr{I}_3
                                                                                                                                    \label{Eq_TRRSPSR_003a}\\
\tsr{b}\tsrbar{S}\tsr{b}+\tsr{b}^2\tsrbar{S}+\tsrbar{S}\tsr{b}^2=&-\II{b}\tsrbar{S}+\I{b\bar{S}}\tsr{b}+\Iab{b}{2}{}{\bar{S}}{}{}\tsr{I}_3
                                                                                                                                    \label{Eq_TRRSPSR_003b}\\
\tsr{b}\tsrbar{S}\tsr{b}^2+\tsr{b}^2\tsrbar{S}\tsr{b}=&\I{b\bar{S}}\tsr{b}^2+\Iab{b}{2}{}{\bar{S}}{}{}\tsr{b}-\III{b}\tsrbar{S}
                                                                                                                                    \label{Eq_TRRSPSR_003c}\\
\tsr{b}\tsrbar{\Omega}\tsr{b}=&-\tsr{b}^2\tsrbar{\Omega}-\tsrbar{\Omega}\tsr{b}^2-\II{b}\tsrbar{\Omega}
                                                                                                                                    \label{Eq_TRRSPSR_003d}\\
\tsr{b}\tsrbar{\Omega}\tsr{b}^2+\tsr{b}^2\tsrbar{\Omega}\tsr{b}=&-\III{b}\tsrbar{\Omega}
                                                                                                                                    \label{Eq_TRRSPSR_003e}
\end{alignat}
\end{subequations}
The most general model for $\phi^{({\rm s};{\mathfrak V})}_{ij}$ \eqref{Eq_TRRSPSR_001} is \cite{Lumley_1978a} a linear combination of $\tsr{b}$ \eqref{Eq_TRRSPSR_002c}
and $\tsr{b}^2+\tfrac{2}{3}\II{b}\tsr{I}_3$ ($\Ia{b}{2}{}=-2\II{b}$) with coeficients which are functions of the invariants $\II{b}$ and $\III{b}$, since by construction $\I{b}=0$ \eqref{Eq_TRRSPSR_002c}.
The most general model for $\phi^{({\rm r};{\mathfrak V})}_{ij}$ \eqref{Eq_TRRSPSR_001} is \cite{Speziale_Sarkar_Gatski_1991a,
                                                                                               Ristorcelli_Lumley_Abid_1995a,
                                                                                               Fu_Wang_1997a}
a linear combination of the symmetric deviatoric tensors constructed from products of $b_{ij}$ \eqref{Eq_TRRSPSR_002c}, $\bar S_{ij}$ \eqref{Eq_TRRSPSR_002a} and $\bar\Omega_{ij}$ \eqref{Eq_TRRSPSR_002b},
which form a representation basis \cite{Spencer_Rivlin_1959a,
                                        Smith_1971a}
of deviatoric-symmetric-tensor-valued isotropic functions of these 3 tensors, omitting, because of the linearity requirement \cite{Lumley_1978a},
elements nonlinear in the mean-velocity gradients ({\em ie} terms containing $\tsrbar{S}^{n_1}\tsrbar{\Omega}^{n_2}$ with $n_1+n_2>1$).
There are 2 different approaches to constructing a basis: polynomial bases \cite{Spencer_Rivlin_1959a} and functional bases \cite{Smith_1971a}.
Polynomial bases \cite{Spencer_Rivlin_1959a} are formed by all products of integer powers of the generating tensors which cannot be represented as a linear combination of the basis-elements using \tsc{ch}-reduction, {\em ie} identities
obtained from the aformentionned Cayley-Hamilton theorem and its extensions \cite{Rivlin_1955a}, so that the representation coefficients are explicitly known polynomial (hence continuous)
functions of the invariants. Functional bases \cite{Smith_1971a} are potentially more compact, because they further reduce the elements of the corresponding polynomial basis
by \tsc{re}-reduction \cite[pp. 380--382]{Rivlin_Ericksen_1955a}, {\em ie} by solving appropriate linear systems \cite{Smith_1971a}.
As a consequence, it can only be asserted \cite{Smith_1971a} that the coefficients for representing a given product between integer powers of the tensors generating the basis
as a combination of the basis-elements are functions (piecewise rational) of the invariants, not necessarily continuous \cite{Smith_1971a},
which are not always explicitly known.
For this reason \cite{Ristorcelli_Lumley_Abid_1995a} a polynomial representation basis is preferable, and in the present case this is \cite{Spencer_Rivlin_1959a,
                                                                                                                                            Ristorcelli_Lumley_Abid_1995a}
$\mathfrak{B}[\tsr{b}, \devnd{S}, \devnd{\Omega}]:=\big\{\devnd{T}_{(n)},\;n\in\{1,\cdots,8\}\big\}$ with
\begin{subequations}
                                                                                                                                    \label{Eq_TRRSPSR_004}
\begin{alignat}{6}
\devnd{T}_{(1)}:=&\tsr{b}
                = b_{ij}\vec{e}_i\otimes\vec{e}_j
                                                                                                                                    \label{Eq_TRRSPSR_004a}\\
\devnd{T}_{(2)}:=&\tsr{b}^2-\tfrac{1}{3}\Ia{b}{2}\;\tsr{I}_3
                = \big(b_{i\ell}b_{\ell j}+\tfrac{2}{3}\II{b}\delta_{ij}\big)\vec{e}_i\otimes\vec{e}_j
                                                                                                                                    \label{Eq_TRRSPSR_004b}\\
\devnd{T}_{(3)}:=&\devnd{S}:= \big(\mathrm{k}\varepsilon^{-1}\big)\bar S_{ij}\vec{e}_i\otimes\vec{e}_j
                                                                                                                                    \label{Eq_TRRSPSR_004c}\\
\devnd{T}_{(4)}:=&\tsr{b}\devnd{S}+\devnd{S}\tsr{b}-\tfrac{2}{3}\I{b\strd{S}}\;\tsr{I}_3
                  \stackrel{\eqrefsabc{Eq_TRRSPSR_002}{Eq_TRRSPSR_004}{Eq_TRRSPSR_005}}{=} -\tfrac{1}{2}\devnd{P}_S-\tfrac{2}{3}\devnd{S}
                                                                                                                                    \label{Eq_TRRSPSR_004d}\\
\devnd{T}_{(5)}:=&\tsr{b}\devnd{\Omega}-\devnd{\Omega}\tsr{b}:=\big(\mathrm{k}\varepsilon^{-1}\big)\big(\tsr{b}\tsrbar{\Omega}-\tsrbar{\Omega}\tsr{b}\big)
                  \stackrel{\eqrefsab{Eq_TRRSPSR_002}{Eq_TRRSPSR_005}}{=}\tfrac{1}{2}\devnd{P}_\Omega
                                                                                                                                    \label{Eq_TRRSPSR_004e}\\
\devnd{T}_{(6)}:=&\tsr{b}^2\devnd{S}+\devnd{S}\tsr{b}^2-\tfrac{2}{3}\Iab{b}{2}{}{\strd{S}}{}{}\;\tsr{I}_3
                                                                                                                                    \label{Eq_TRRSPSR_004f}\\
\devnd{T}_{(7)}:=&\tsr{b}^2\devnd{\Omega}-\devnd{\Omega}\tsr{b}^2
                                                                                                                                    \label{Eq_TRRSPSR_004g}\\
\devnd{T}_{(8)}:=&\tsr{b}^2\devnd{\Omega}\tsr{b}-\tsr{b}\devnd{\Omega}\tsr{b}^2
                                                                                                                                    \label{Eq_TRRSPSR_004h}
\end{alignat}
\end{subequations}
where the 8 tensors $\devnd{T}_{(n)}$ in \eqref{Eq_TRRSPSR_004} were made nondimensional 
by scaling with appropriate powers of the turbulence kinetic energy $\mathrm{k}:=\tfrac{1}{2}\overline{u'_\ell u'_\ell}$
and its dissipation-rate $\varepsilon:=2\nu\overline{\partial_{x_j} u'_i\partial_{x_j} u'_i}$.
The above basis, $\mathfrak{B}[\tsr{b}, \devnd{S}, \devnd{\Omega}]:=\big\{\devnd{T}_{(n)},\;n\in\{1,\cdots,8\}\big\}$ \eqref{Eq_TRRSPSR_004},
is the classical \cite{Ristorcelli_Lumley_Abid_1995a} representation basis in homogeneous turbulence, where $\tsrbar{S}$ and $\tsrbar{\Omega}$ are usually fixed inputs of the problem \cite{Reynolds_1974a,
                                                                                                                                                                                             Lumley_1978a}.

Notice \cite{Ristorcelli_Lumley_Abid_1995a} that $\devnd{T}_{(8)}$ \eqref{Eq_TRRSPSR_004h} can in principle be projected by \tsc{re}-reduction to obtain an irreducible
functional representation basis \cite{Speziale_Sarkar_Gatski_1991a}, $\mathfrak{B}[\tsr{b}, \devnd{S}, \devnd{\Omega}]\setminus\{\devnd{T}_{(8)}\}$,
but the coefficients for this projection are not known explicitly. Furthermore, even if the projection were to be sought, on a value-by-value basis \cite[pp. 380--382]{Rivlin_Ericksen_1955a},
the resulting representation coefficients would not necessarily be continuous functions of the invariants \cite{Smith_1971a},
and this makes functional representation bases awkward to use. The problem is even more acute in representation bases for wall turbulence \cite{Gerolymos_Sauret_Vallet_2004a},
where the basic Gibson-Launder \cite{Gibson_Launder_1978a} rapid wall-echo model cannot be explicitly represented in the functional basis built by $[\tsr{b}, \devnd{S}, \devnd{\Omega}]$ and the unit-vector in
the normal-to-the-wall direction $\vec{e}_n$. Therefore, as noted in \cite{Ristorcelli_Lumley_Abid_1995a}, polynomial representation bases are the best choice, because the increased number of basis-elements
is justified by the possibility of explicit continuous representation of models, using the Cayley-Hamilton theorem and its extensions \cite{Rivlin_1955a}.
\begin{table}[ht]
\begin{center}
\scalebox{0.8}{
\begin{tabular}{lcccccccc}
$\displaystyle\genfrac{}{}{0pt}{}{}{n}\Big\backslash\genfrac{}{}{0pt}{}{m}{}$
          &                         $1$&                     $2$&                     $3$&                   $4$&                 $5$&                $6$&              $7$&            $8$\\\\\hline
       $1$&$1$                         &$0$                     &$0$                     &$0$                   &$0$                 &$0$                &$0$              &$0$            \\
       $2$&$0$                         &$1$                     &$0$                     &$0$                   &$0$                 &$0$                &$0$              &$0$            \\
       $3$&$0$                         &$0$                     &$1$                     &$0$                   &$0$                 &$0$                &$0$              &$0$            \\
       $4$&$0$                         &$0$                     &$-\tfrac{4}{3}$         &$-2$                  &$0$                 &$0$                &$0$              &$0$            \\
       $5$&$0$                         &$0$                     &$0$                     &$0$                   &$2$                 &$0$                &$0$              &$0$            \\
       $6$&$-\tfrac{4}{3}\I{b\strd{S}}$&$0$                     &$4\II{b}$               &$-\tfrac{4}{3}$       &$0$                 &$2$                &$0$              &$0$            \\
       $7$&$0$                         &$0$                     &$0$                     &$0$                   &$0$                 &$0$                &$2$              &$0$            \\
       $8$&$0$                         &$0$                     &$0$                     &$0$                   &$-2\II{b}$          &$0$                &$0$              &$-2$           \\
\end{tabular}
}\\
\caption{Matrix of coefficients $a_{\tsc{ht}_{nm}}$ for the representation $\devnd{H}_{(n)}=\sum_{m=1}^{8}a_{\tsc{ht}_{nm}}\devnd{T}_{(m)}$ \eqref{Eq_TRRSPSR_007}
         of the elements $\devnd{H}_{(n)}\in\mathfrak{B}_{(P)}[\tsr{b}, \devnd{S}, \devnd{P}_{\bar\Omega}; \devnd{P}_{\bar S}]$ \eqref{Eq_TRRSPSR_005}
         as a linear combination of the elements $\devnd{T}_{(m)}\in\mathfrak{B}[\tsr{b}, \devnd{S}, \devnd{\Omega}]$ \eqref{Eq_TRRSPSR_004}.}
\label{Tab_TRRSPSR_001}
\end{center}
%
\begin{center}
\scalebox{0.8}{
\begin{tabular}{lcccccccc}
$\displaystyle\genfrac{}{}{0pt}{}{}{n}\Big\backslash\genfrac{}{}{0pt}{}{m}{}$
          &                         $1$&                     $2$&                     $3$&                   $4$&                 $5$&                $6$&              $7$&            $8$\\\\\hline
       $1$&$1$                         &$0$                     &$0$                     &$0$                   &$0$                 &$0$                &$0$              &$0$            \\
       $2$&$0$                         &$1$                     &$0$                     &$0$                   &$0$                 &$0$                &$0$              &$0$            \\
       $3$&$0$                         &$0$                     &$1$                     &$0$                   &$0$                 &$0$                &$0$              &$0$            \\
       $4$&$0$                         &$0$                     &$-\tfrac{2}{3}$         &$-\tfrac{1}{2}$       &$0$                 &$0$                &$0$              &$0$            \\
       $5$&$0$                         &$0$                     &$0$                     &$0$                   &$\tfrac{1}{2}$      &$0$                &$0$              &$0$            \\
       $6$&$-\tfrac{1}{3}\dfrac{P_\mathrm{k}}{\rho\varepsilon}$
                                       &$0$                     &$-2\II{b}-\tfrac{4}{9}$ &$-\tfrac{1}{3}$       &$0$                 &$\tfrac{1}{2}$     &$0$              &$0$            \\
       $7$&$0$                         &$0$                     &$0$                     &$0$                   &$0$                 &$0$                &$\tfrac{1}{2}$   &$0$            \\
       $8$&$0$                         &$0$                     &$0$                     &$0$                   &$-\tfrac{1}{2}\II{b}$
                                                                                                                                     &$0$                &$0$              &$-\tfrac{1}{2}$\\
\end{tabular}
}\\
\caption{Matrix of coefficients $a_{{\tsc{th}}_{nm}}$ for the representation $\devnd{T}_{(n)}=\sum_{m=1}^{8}a_{\tsc{th}_{nm}}\devnd{H}_{(m)}$ \eqref{Eq_TRRSPSR_007}
         of the elements $\devnd{T}_{(n)}\in\mathfrak{B}[\tsr{b}, \devnd{S}, \devnd{\Omega}]$ \eqref{Eq_TRRSPSR_004}
         as a linear combination of the elements $\devnd{H}_{(m)}\in\mathfrak{B}_{(P)}[\tsr{b}, \devnd{S}, \devnd{P}_{\bar\Omega}; \devnd{P}_{\bar S}]$ \eqref{Eq_TRRSPSR_005}.}
\label{Tab_TRRSPSR_002}
\end{center}
\end{table}

%
%
%
%
%
\section{Production by strain or rotation}\label{TRRSPSR_s_PSR}
%
%
%
%
%

The tensor $\devnd{T}_{(4)}$ is related to strain-production \eqref{Eq_TRRSPSR_004d} and the tensor $\devnd{T}_{(5)}$ to rotation-production \eqref{Eq_TRRSPSR_004e}. We can therefore construct
an equivalent representation basis  
$\mathfrak{B}_{(P)}[\tsr{b}, \devnd{S}, \devnd{P}_{\bar\Omega}; \devnd{P}_{\bar S}]:=\big\{\devnd{H}_{(n)},\;n\in\{1,\cdots,8\}\big\}$ with
\begin{subequations}
                                                                                                                                    \label{Eq_TRRSPSR_005}
\begin{alignat}{6}
\devnd{H}_{(1)}:=&\tsr{b}
                  \stackrel{\eqref{Eq_TRRSPSR_004a}}{=:} \devnd{T}_{(1)}
                                                                                                                                    \label{Eq_TRRSPSR_005a}\\
\devnd{H}_{(2)}:=&\tsr{b}^2-\tfrac{1}{3}\Ia{b}{2}\;\tsr{I}_3
                  \stackrel{\eqref{Eq_TRRSPSR_004b}}{=:} \devnd{T}_{(2)}
                                                                                                                                    \label{Eq_TRRSPSR_005b}\\
\devnd{H}_{(3)}:=&\devnd{S}
                  \stackrel{\eqref{Eq_TRRSPSR_004c}}{=:} \devnd{T}_{(3)}
                                                                                                                                    \label{Eq_TRRSPSR_005c}\\
\devnd{H}_{(4)}:=&\devnd{P}_{\bar S}:=\big(\rho\varepsilon\big)^{-1}\big(P_{\bar S_{ij}}-\tfrac{2}{3}P_\mathrm{k}\delta_{ij}\big)\vec{e}_i\otimes\vec{e}_j
                                                                                                                                    \notag\\
                 \stackrel{\eqrefsab{Eq_TRRSPSR_004c}{Eq_TRRSPSR_004d}}{=}&-2\devnd{T}_{(4)}-\tfrac{4}{3}\devnd{T}_{(3)}
                                                                                                                                    \label{Eq_TRRSPSR_005d}\\
\devnd{H}_{(5)}:=&\devnd{P}_{\bar\Omega}:=\big(\rho\varepsilon\big)^{-1}P_{\bar\Omega_{ij}}\vec{e}_i\otimes\vec{e}_j
                 \stackrel{\eqref{Eq_TRRSPSR_004e}}{=}2\devnd{T}_{(5)}
                                                                                                                                    \label{Eq_TRRSPSR_005e}\\
\devnd{H}_{(6)}:=&\tsr{b}\devnd{P}_{\bar S}+\devnd{P}_{\bar S}\tsr{b}-\tfrac{2}{3}\Iab{b}{}{}{\strd{P}}{}{\bar S}\;\tsr{I}_3
                                                                                                                                    \notag\\
                 \stackrel{\eqrefsabc{Eq_TRRSPSR_002}{Eq_TRRSPSR_004}{Eq_TRRSPSR_003b}}{=}& -\tfrac{4}{3}\I{b\devnd{S}}\devnd{T}_{(1)}
                                                                                      +4\II{b}                   \devnd{T}_{(3)}
                                                                                      -\tfrac{4}{3}              \devnd{T}_{(4)}
                                                                                      +2                         \devnd{T}_{(6)}
                                                                                                                                    \label{Eq_TRRSPSR_005f}\\
\devnd{H}_{(7)}:=&\tsr{b}\devnd{P}_{\bar\Omega}+\devnd{P}_{\bar\Omega}\tsr{b}
                 \stackrel{\eqrefsabc{Eq_TRRSPSR_002}{Eq_TRRSPSR_004g}{Eq_TRRSPSR_003d}}{=}2\devnd{T}_{(7)}
                                                                                                                                    \label{Eq_TRRSPSR_005g}\\
\devnd{H}_{(8)}:=&\tsr{b}^2\devnd{P}_{\bar\Omega}+\devnd{P}_{\bar\Omega}\tsr{b}^2
                 \stackrel{\eqrefsabc{Eq_TRRSPSR_002}{Eq_TRRSPSR_004}{Eq_TRRSPSR_003a}}{=} -2\II{b}                                              \devnd{T}_{(5)}
                                                                                    -2                                                    \devnd{T}_{(8)}
                                                                                                                                    \label{Eq_TRRSPSR_005h}
\end{alignat}
\end{subequations}
where the identities in \eqref{Eq_TRRSPSR_005} are
obtained by direct computation from definitions \eqrefsatob{Eq_TRRSPSR_002c}{Eq_TRRSPSR_002e},
using the following relations between invariants,
also obtained by direct computation from definitions \eqref{Eq_TRRSPSR_002}
\begin{subequations}
                                                                                                                                    \label{Eq_TRRSPSR_006}
\begin{alignat}{6}
\Ia{\strd{P}}{}{\bar\Omega}\stackrel{\eqref{Eq_TRRSPSR_002}}{=}&\Iab{b}{}{}{\strd{P}}{}{\bar\Omega}\stackrel{\eqref{Eq_TRRSPSR_002}}{=}\Iab{b}{2}{}{\strd{P}}{}{\bar\Omega}\stackrel{\eqref{Eq_TRRSPSR_002}}{=}0
                                                                                                                                    \label{Eq_TRRSPSR_006a}\\
2P_\mathrm{k}\stackrel{\eqref{Eq_TRRSPSR_002f}}{=:}&                    \I{P}
             \stackrel{\eqrefsab{Eq_TRRSPSR_002d}{Eq_TRRSPSR_002f}}{=}\Ia{P}{}{\bar S}
             \stackrel{\eqrefsab{Eq_TRRSPSR_002g}{Eq_TRRSPSR_006a}}{=}\I{D}
                                                                                                                                    \label{Eq_TRRSPSR_006b}\\
\I{b\strd{S}}\stackrel{\eqrefsab{Eq_TRRSPSR_004d}{Eq_TRRSPSR_006b}}{=}&-\tfrac{1}{2}\dfrac{P_\mathrm{k}}{\rho\varepsilon}
                                                                                                                                    \label{Eq_TRRSPSR_006c}\\
\I{b\strd{D}}          \stackrel{\eqrefsabc{Eq_TRRSPSR_008c}{Eq_TRRSPSR_008d}{Eq_TRRSPSR_002}}{=}&
\I{b\strd{P}}          \stackrel{\eqrefsabc{Eq_TRRSPSR_005d}{Eq_TRRSPSR_008c}{Eq_TRRSPSR_002}}{=}
\Ia{b\strd{P}}{}{\bar S}\stackrel{\eqrefsabc{Eq_TRRSPSR_004c}{Eq_TRRSPSR_005d}{Eq_TRRSPSR_002}}{=}-4\Iab{b}{2}{}{\strd{S}}{}{}-\tfrac{4}{3}\I{b\strd{S}}
                                                                                                                                    \label{Eq_TRRSPSR_006d}\\
\Iab{b}{2}{}{\strd{S}}{}{}\stackrel{\eqrefsab{Eq_TRRSPSR_006d}{Eq_TRRSPSR_006c}}{=}&\tfrac{1}{6}\dfrac{P_\mathrm{k}}{\rho\varepsilon}-\tfrac{1}{4}\Iab{b}{}{}{\strd{P}}{}{\bar S}
                                                                                                                                    \label{Eq_TRRSPSR_006e}\\
\Iab{b}{2}{}{\strd{P}}{}{}\stackrel{\eqrefsabc{Eq_TRRSPSR_002}{Eq_TRRSPSR_008}{Eq_TRRSPSR_006}}{=}&\Iab{b}{2}{}{\strd{D}}{}{}
                          \stackrel{\eqrefsabc{Eq_TRRSPSR_002}{Eq_TRRSPSR_005}{Eq_TRRSPSR_008}}{=}\Iab{b}{2}{}{\strd{P}}{}{\bar S}
                                                                                                                                    \notag\\
                          \stackrel{\eqrefsabc{Eq_TRRSPSR_002}{Eq_TRRSPSR_005}{Eq_TRRSPSR_006}}{=}&
-\big(\tfrac{2}{9}+\tfrac{2}{3}\II{b}\big)\dfrac{P_\mathrm{k}}{\rho\varepsilon}+\tfrac{1}{3}\Iab{b}{}{}{\strd{P}}{}{\bar S}
                                                                                                                                    \label{Eq_TRRSPSR_006f}
\end{alignat}
\end{subequations}
From the relations \eqref{Eq_TRRSPSR_006} we can readily identify the matrix of coefficients $\underline{\underline{a}}_{(\tsc{ht})}:=[a_{{(\tsc{ht})}_{nm}}]\in\mathbb{R}^{8\times8}$ \tabref{Tab_TRRSPSR_001},
and calculate its inverse $\underline{\underline{a}}_{(\tsc{th})}:=\underline{\underline{a}}_{\tsc{ht}}^{-1}$ \tabref{Tab_TRRSPSR_002}, which relate the column-vectors of basis-elements
$\underline{\devnd{T}}:=[\devnd{T}_{(1)},\cdots,\devnd{T}_{(8)}]^\tsc{T}$ \eqref{Eq_TRRSPSR_004} and $\underline{\devnd{H}}:=[\devnd{H}_{(1)},\cdots,\devnd{H}_{(8)}]^\tsc{T}$ \eqref{Eq_TRRSPSR_005}
\begin{alignat}{6}
\underline{\devnd{H}}=\underline{\underline{a}}_{\tsc{ht}}\;\underline{\devnd{T}}
\iff
\underline{\devnd{T}}=\underbrace{\underline{\underline{a}}_{\tsc{ht}}^{-1}}_{\displaystyle \underline{\underline{a}}_{(\tsc{th})}}\underline{\devnd{H}}
                                                                                                                                    \label{Eq_TRRSPSR_007}
\end{alignat}
The passage-matrix $\underline{\underline{a}}_{(\tsc{ht})}$ \tabref{Tab_TRRSPSR_001}, and its inverse $\underline{\underline{a}}_{(\tsc{th})}$ \tabref{Tab_TRRSPSR_002},
can be used to systematically project models from one basis to another (\S\ref{TRRSPSR_s_CoBR}).

%
%
%
%
%
\section{Symmetric representation bases}\label{TRRSPSR_s_SRB}
%
%
%
%
%

Many early models \cite{Reynolds_1974a,
                        Naot_Shavit_Wolfshtein_1973a,
                        Launder_Reece_Rodi_1975a}
used the rate-of-production tensor $P_{ij}$ \eqref{Eq_TRRSPSR_002f} and the tensor $D_{ij}$ which we identified in \eqref{Eq_TRRSPSR_002g} as the difference
between strain-production minus rotation-production. These 2 tensors are used by several researchers \cite{Fu_1988a,
                                                                                                           Craft_Launder_2001a},
who sometimes \cite{So_Aksoy_Yuan_Sommer_1996a} express models developed in the classical basis $\mathfrak{B}[\tsr{b}, \devnd{S}, \devnd{\Omega}]$ \eqref{Eq_TRRSPSR_004}
in terms of $P_{ij}$ \eqref{Eq_TRRSPSR_002f} and $D_{ij}$ \eqref{Eq_TRRSPSR_002g}.
If we consider the tensors $P_{ij}$ \eqref{Eq_TRRSPSR_002f} and $D_{ij}$ \eqref{Eq_TRRSPSR_002g} representative of the influence of mean-velocity gradients $\partial_{x_j}\bar u_i$ on
the $\phi_{ij}$ \eqref{Eq_TRRSPSR_001}, since it is $P_{ij}$ \eqref{Eq_TRRSPSR_002f}, and not $\bar S_{ij}$ \eqref{Eq_TRRSPSR_002a} alone,
which appears directly in the transport equations for the Reynolds-stresses \cite{Speziale_Sarkar_Gatski_1991a},
we may construct a representation basis generated by $b_{ij}$ \eqref{Eq_TRRSPSR_002c} and these 2 tensors,
$\mathfrak{B}[\tsr{b}, \devnd{P}, \devnd{D}]:=\big\{\devnd{J}_{(n)},\;n\in\{1,\cdots,8\}\big\}$ with
\begin{subequations}
                                                                                                                                    \label{Eq_TRRSPSR_008}
\begin{alignat}{6}
\devnd{J}_{(1)}:=&\tsr{b}
                   \stackrel{\eqref{Eq_TRRSPSR_002a}}{=:} \devnd{T}_{(1)}
                                                                                                                                    \label{Eq_TRRSPSR_008a}\\
\devnd{J}_{(2)}:=&\tsr{b}^2-\tfrac{1}{3}\Ia{b}{2}\;\tsr{I}_3
                   \stackrel{\eqref{Eq_TRRSPSR_002b}}{=:} \devnd{T}_{(2)}
                                                                                                                                    \label{Eq_TRRSPSR_008b}
\end{alignat}
\begin{alignat}{6}
\devnd{J}_{(3)}:=&
\devnd{P}:=\big(\rho\varepsilon\big)^{-1}\big(P_{ij}-\tfrac{2}{3}P_\mathrm{k}\delta_{ij}\big)\vec{e}_i\otimes\vec{e}_j\stackrel{\eqrefsabc{Eq_TRRSPSR_002f}{Eq_TRRSPSR_005d}{Eq_TRRSPSR_005e}}{=}
                                                                                                                                    \notag\\
                                                                          &-\tfrac{4}{3}\devnd{T}_{(3)}
                                                                            -2           \devnd{T}_{(4)}
                                                                            +2           \devnd{T}_{(5)}
                                                                                                                                    \label{Eq_TRRSPSR_008c}\\
\devnd{J}_{(4)}:=&\devnd{D}:=\big(\rho\varepsilon\big)^{-1}\big(D_{ij}-\tfrac{2}{3}P_\mathrm{k}\delta_{ij}\big)\vec{e}_i\otimes\vec{e}_j\stackrel{\eqrefsabc{Eq_TRRSPSR_002g}{Eq_TRRSPSR_005d}{Eq_TRRSPSR_005e}}{=}
                                                                                                                                    \notag\\
                                                                           &-\tfrac{4}{3}\devnd{T}_{(3)}
                                                                            -2           \devnd{T}_{(4)}
                                                                            -2           \devnd{T}_{(5)}
                                                                                                                                    \label{Eq_TRRSPSR_008d}\\
\devnd{J}_{(5)}:=&
\tsr{b}\devnd{P}+\devnd{P}\tsr{b}-\tfrac{ 2}{3}\I{b\devnd{P}}\tsr{I}_3\stackrel{\eqrefsabc{Eq_TRRSPSR_002f}{Eq_TRRSPSR_005f}{Eq_TRRSPSR_005g}}{=}
                                                                                                                                    \notag\\
                                                                           &-\tfrac{4}{3}\I{b\devnd{S}}\devnd{T}_{(1)}
                                                                            +4\II{b}                   \devnd{T}_{(3)}
                                                                            -\tfrac{4}{3}              \devnd{T}_{(4)}
                                                                            +2                         \devnd{T}_{(6)}
                                                                            +2                         \devnd{T}_{(7)}
                                                                                                                                    \label{Eq_TRRSPSR_008e}\\
\devnd{J}_{(6)}:=&
\tsr{b}\devnd{D}+\devnd{D}\tsr{b}-\tfrac{ 2}{3}\I{b\devnd{D}}\tsr{I}_3\stackrel{\eqrefsabc{Eq_TRRSPSR_002g}{Eq_TRRSPSR_005f}{Eq_TRRSPSR_005g}}{=}
                                                                                                                                    \notag\\
                                                                           &-\tfrac{4}{3}\I{b\devnd{S}}\devnd{T}_{(1)}
                                                                            +4\II{b}                   \devnd{T}_{(3)}
                                                                            -\tfrac{4}{3}              \devnd{T}_{(4)}
                                                                            +2                         \devnd{T}_{(6)}
                                                                            -2                         \devnd{T}_{(7)}
                                                                                                                                    \label{Eq_TRRSPSR_008f}\\
\devnd{J}_{(7)}:=&
\tsr{b}^2\devnd{P}+\devnd{P}\tsr{b}^2-\tfrac{ 2}{3}\Iab{b}{2}{}{\devnd{P}}{}{}\tsr{I}_3\stackrel{\eqrefsabc{Eq_TRRSPSR_002f}{Eq_TRRSPSR_005h}{Eq_TRRSPSR_009g}}{=}
                                                                                                                                    \notag\\
                                                                                    &-2\Iab{b}{2}{}{\devnd{S}}{}{}\devnd{T}_{(1)}
                                                                                     +\tfrac{2}{3}\I{b\devnd{S}}  \devnd{T}_{(2)}
                                                                                     -2\III{b}                    \devnd{T}_{(3)}
                                                                                                                                    \notag\\
                                                                                    &+2\II{b}                     \devnd{T}_{(4)}
                                                                                     -2\II{b}                     \devnd{T}_{(5)}
                                                                                     -\tfrac{4}{3}                \devnd{T}_{(6)}
                                                                                     -2                           \devnd{T}_{(8)}
                                                                                                                                    \label{Eq_TRRSPSR_008g}\\
\devnd{J}_{(8)}:=&
\tsr{b}^2\devnd{D}+\devnd{D}\tsr{b}^2-\tfrac{ 2}{3}\Iab{b}{2}{}{\devnd{P}}{}{}\tsr{I}_3\stackrel{\eqrefsabc{Eq_TRRSPSR_002g}{Eq_TRRSPSR_005h}{Eq_TRRSPSR_009g}}{=}
                                                                                                                                    \notag\\
                                                                                       &-2\Iab{b}{2}{}{\devnd{S}}{}{}                             \devnd{T}_{(1)}
                                                                                         +\tfrac{2}{3}\I{b\devnd{S}}                           \devnd{T}_{(2)}
                                                                                         -2\III{b}                                             \devnd{T}_{(3)}
                                                                                                                                    \notag\\
                                                                                        &+2\II{b}                                              \devnd{T}_{(4)}
                                                                                         +2\II{b}                                              \devnd{T}_{(5)}
                                                                                         -\tfrac{4}{3}                                         \devnd{T}_{(6)}
                                                                                         +2                                                    \devnd{T}_{(8)}
                                                                                                                                    \label{Eq_TRRSPSR_008h}
\end{alignat}
\end{subequations}
where the relations with the elements of $\mathfrak{B}[\tsr{b}, \devnd{S}, \devnd{\Omega}]:=\big\{\devnd{T}_{(n)},\;n\in\{1,\cdots,8\}\big\}$ \eqref{Eq_TRRSPSR_004}
are obtained by direct computation using the expressions \eqref{Eq_TRRSPSR_006} for the invariants, and the aformentionned Cayley-Hamilton theorem and its extensions \cite{Rivlin_1955a}.
Alternatively we can use
$P_{\bar S_{ij}}\stackrel{\eqrefsab{Eq_TRRSPSR_002f}{Eq_TRRSPSR_002g}}{=}\tfrac{1}{2}(P_{ij}+D_{ij})$
and
$P_{\bar\Omega_{ij}}\stackrel{\eqrefsab{Eq_TRRSPSR_002f}{Eq_TRRSPSR_002g}}{=}\tfrac{1}{2}(P_{ij}-D_{ij})$
to obtain an equivalent representation basis
generated by the 3 symmetric tensors $\tsr{b}$, $\devnd{P}_{\bar S}$ and $\devnd{P}_{\bar \Omega}$,
$\mathfrak{B}[\tsr{b}, \devnd{P}_{\bar S}, \devnd{P}_{\bar\Omega}]:=\big\{\devnd{F}_{(n)},\;n\in\{1,\cdots,8\}\big\}$ with
\begin{subequations}
                                                                                                                                    \label{Eq_TRRSPSR_009}
\begin{alignat}{6}
\devnd{F}_{(1)}:=&\tsr{b}
                   \stackrel{\eqref{Eq_TRRSPSR_002a}}{=:} \devnd{T}_{(1)}
                                                                                                                                    \label{Eq_TRRSPSR_009a}\\
\devnd{F}_{(2)}:=&\tsr{b}^2-\tfrac{1}{3}\Ia{b}{2}\;\tsr{I}_3
                   \stackrel{\eqref{Eq_TRRSPSR_002b}}{=:} \devnd{T}_{(2)}
                                                                                                                                    \label{Eq_TRRSPSR_009b}\\
\devnd{F}_{(3)}:=&\devnd{P}_{\bar S}\stackrel{\eqref{Eq_TRRSPSR_005d}}{=:}\devnd{H}_{(4)}
                                                                                                                                    \label{Eq_TRRSPSR_009c}\\
\devnd{F}_{(4)}:=&\devnd{P}_{\bar\Omega}\stackrel{\eqref{Eq_TRRSPSR_005e}}{=:}\devnd{H}_{(5)}
                                                                                                                                    \label{Eq_TRRSPSR_009d}\\
\devnd{F}_{(5)}:=&\tsr{b}\devnd{P}_{\bar S}+\devnd{P}_{\bar S}\tsr{b}-\tfrac{2}{3}\Iab{b}{}{}{\strd{P}}{}{\bar S}\;\tsr{I}_3\stackrel{\eqref{Eq_TRRSPSR_005f}}{=:}\devnd{H}_{(6)}
                                                                                                                                    \label{Eq_TRRSPSR_009e}\\
\devnd{F}_{(6)}:=&\tsr{b}\devnd{P}_{\bar\Omega}+\devnd{P}_{\bar\Omega}\tsr{b}\stackrel{\eqref{Eq_TRRSPSR_005g}}{=:}\devnd{H}_{(7)}
                                                                                                                                    \label{Eq_TRRSPSR_009f}\\
\devnd{F}_{(7)}:=&\tsr{b}^2\devnd{P}_{\bar S}+\devnd{P}_{\bar S}\tsr{b}^2-\tfrac{2}{3}\Iab{b}{2}{}{\strd{P}}{}{\bar S}\;\tsr{I}_3
                                                                                                                                    \notag\\
                \stackrel{\eqrefsabc{Eq_TRRSPSR_003a}{Eq_TRRSPSR_003c}{Eq_TRRSPSR_004}}{=}&-2\Iab{b}{2}{}{\strd{S}}{}{}\devnd{T}_{(1)}
                                                                                           +\tfrac{2}{3}\I{b\strd{S}}  \devnd{T}_{(2)}
                                                                                                                                    \notag\\
                                                                                          &-2\III{b}                   \devnd{T}_{(3)}
                                                                                           +2\II{b}                    \devnd{T}_{(4)}
                                                                                           -\tfrac{4}{3}               \devnd{T}_{(6)}
                                                                                                                                    \label{Eq_TRRSPSR_009g}\\
\devnd{F}_{(8)}:=&\tsr{b}^2\devnd{P}_{\bar\Omega}+\devnd{P}_{\bar\Omega}\tsr{b}^2\stackrel{\eqref{Eq_TRRSPSR_005h}}{=:}\devnd{H}_{(8)}
                                                                                                                                    \label{Eq_TRRSPSR_009h}
\end{alignat}
\end{subequations}
\begin{table}[ht]
\begin{center}
\scalebox{0.8}{
\begin{tabular}{lcccccccc}
$\displaystyle\genfrac{}{}{0pt}{}{}{n}\Big\backslash\genfrac{}{}{0pt}{}{m}{}$
          &                         $1$&                     $2$&                     $3$&                   $4$&                 $5$&                $6$&              $7$&            $8$\\\\\hline
       $1$&$1$                         &$0$                     &$0$                     &$0$                   &$0$                 &$0$                &$0$              &$0$            \\
       $2$&$0$                         &$1$                     &$0$                     &$0$                   &$0$                 &$0$                &$0$              &$0$            \\
       $3$&$0$                         &$0$                     &$-\tfrac{4}{3}$         &$-2$                  &$+2$                &$0$                &$0$              &$0$            \\
       $4$&$0$                         &$0$                     &$-\tfrac{4}{3}$         &$-2$                  &$-2$                &$0$                &$0$              &$0$            \\
       $5$&$-\tfrac{4}{3}\I{b\strd{S}}$&$0$                     &$4\II{b}$               &$-\tfrac{4}{3}$       &$0$                 &$+2$               &$+2$             &$0$            \\
       $6$&$-\tfrac{4}{3}\I{b\strd{S}}$&$0$                     &$4\II{b}$               &$-\tfrac{4}{3}$       &$0$                 &$+2$               &$-2$             &$0$            \\
       $7$&$-2\Iab{b}{2}{}{\strd{S}}{}{}$
                                       &$\tfrac{2}{3}\I{b\strd{S}}$
                                                                &$-2\III{b}$             &$+2\II{b}$            &$-2\II{b}$          &$-\tfrac{4}{3}$    &$0$              &$-2$           \\
       $8$&$-2\Iab{b}{2}{}{\strd{S}}{}{}$
                                       &$\tfrac{2}{3}\I{b\strd{S}}$
                                                                &$-2\III{b}$             &$+2\II{b}$            &$+2\II{b}$          &$-\tfrac{4}{3}$    &$0$              &$+2$           \\
\end{tabular}
}\\
\caption{Matrix of coefficients $a_{{\tsc{jt}}_{nm}}$ for the representation $\devnd{J}_{(n)}=\sum_{m=1}^{8}a_{\tsc{jt}_{nm}}\devnd{T}_{(m)}$ \eqref{Eq_TRRSPSR_010a}
         of the elements $\devnd{J}_{(n)}\in\mathfrak{B}[\tsr{b}, \devnd{P}, \devnd{D}]$ \eqref{Eq_TRRSPSR_008}
         as a linear combination of the elements $\devnd{T}_{(m)}\in\mathfrak{B}[\tsr{b}, \devnd{S}, \devnd{\Omega}]$ \eqref{Eq_TRRSPSR_004}.}
\label{Tab_TRRSPSR_003}
\end{center}
\begin{center}
\scalebox{0.8}{
\begin{tabular}{lcccccccc}
$\displaystyle\genfrac{}{}{0pt}{}{}{n}\Big\backslash\genfrac{}{}{0pt}{}{m}{}$
          &                         $1$&                     $2$&                     $3$&                   $4$&                 $5$&                $6$&              $7$&            $8$\\\\\hline
       $1$&$1$                         &$0$                     &$0$                     &$0$                   &$0$                 &$0$                &$0$              &$0$            \\
       $2$&$0$                         &$1$                     &$0$                     &$0$                   &$0$                 &$0$                &$0$              &$0$            \\
       $3$&$0$                         &$0$                     &$-\tfrac{4}{3}$         &$-2$                  &$0$                 &$0$                &$0$              &$0$            \\
       $4$&$0$                         &$0$                     &$0$                     &$0$                   &$2$                 &$0$                &$0$              &$0$            \\
       $5$&$-\tfrac{4}{3}\I{b\strd{S}}$&$0$                     &$4\II{b}$               &$-\tfrac{4}{3}$       &$0$                 &$2$                &$0$              &$0$            \\
       $6$&$0$                         &$0$                     &$0$                     &$0$                   &$0$                 &$0$                &$2$              &$0$            \\
       $7$&$-2\Iab{b}{2}{}{\strd{S}}{}{}$
                                       &$\tfrac{2}{3}\I{b\strd{S}}$
                                                                &$-2\III{b}$             &$2\II{b}$             &$0$                 &$-\tfrac{4}{3}$    &$0$              &$0$            \\
       $8$&$0$                         &$0$                     &$0$                     &$0$                   &$-2\II{b}$          &$0$                &$0$              &$-2$           \\
\end{tabular}
}\\
\caption{Matrix of coefficients $a_{{\tsc{ft}}_{nm}}$ for the representation $\devnd{F}_{(n)}=\sum_{m=1}^{8}a_{\tsc{ft}_{nm}}\devnd{T}_{(m)}$ \eqref{Eq_TRRSPSR_010b}
         of the elements $\devnd{F}_{(n)}\in\mathfrak{B}[\tsr{b}, \devnd{P}_{\bar S}, \devnd{P}_{\bar\Omega}]$ \eqref{Eq_TRRSPSR_009}
         as a linear combination of the elements $\devnd{T}_{(m)}\in\mathfrak{B}[\tsr{b}, \devnd{S}, \devnd{\Omega}]$ \eqref{Eq_TRRSPSR_004}.}
\label{Tab_TRRSPSR_004}
\end{center}
\end{table}
where \eqref{Eq_TRRSPSR_009g} is again obtained by direct computation, using \eqrefsabcd{Eq_TRRSPSR_002}
                                                                                        {Eq_TRRSPSR_004}
                                                                                        {Eq_TRRSPSR_005}
                                                                                        {Eq_TRRSPSR_006}.

Each of the bases $\mathfrak{B}[\tsr{b}, \devnd{P}, \devnd{D}]$ \eqref{Eq_TRRSPSR_008} or $\mathfrak{B}[\tsr{b}, \devnd{P}_{\bar S}, \devnd{P}_{\bar\Omega}]$ \eqref{Eq_TRRSPSR_009}
is generated by 3 symmetric deviatoric tensors. Therefore, under the constraint of linearity in mean-velocity gradients \cite{Lumley_1978a},
$\mathfrak{B}[\tsr{b}, \devnd{P}, \devnd{D}]$ \eqref{Eq_TRRSPSR_008} is symmetric in $\devnd{P}$ \eqrefsab{Eq_TRRSPSR_008c}{Eq_TRRSPSR_002f} and $\devnd{D}$ \eqrefsab{Eq_TRRSPSR_008d}{Eq_TRRSPSR_002g}, and
$\mathfrak{B}[\tsr{b}, \devnd{P}_{\bar S}, \devnd{P}_{\bar\Omega}]$ \eqref{Eq_TRRSPSR_009} is symmetric in
$\devnd{P}_{\bar S}$ \eqrefsab{Eq_TRRSPSR_005d}{Eq_TRRSPSR_002d} and $\devnd{P}_{\bar\Omega}$ \eqrefsab{Eq_TRRSPSR_005e}{Eq_TRRSPSR_002e}.
On the contrary, the classical representation basis $\mathfrak{B}[\tsr{b}, \devnd{S}, \devnd{\Omega}]$ \eqref{Eq_TRRSPSR_004},
being built by the 2 symmetric tensors $\tsr{b}$ \eqref{Eq_TRRSPSR_002c} and $\tsrbar{S}$ \eqref{Eq_TRRSPSR_002a}, and the antisymmetric tensor $\tsrbar{\Omega}$ \eqref{Eq_TRRSPSR_002b},
contains the element $\devnd{T}_{(8)}:=\tsr{b}^2\devnd{\Omega}\tsr{b}-\tsr{b}\devnd{\Omega}\tsr{b}^2$ \eqref{Eq_TRRSPSR_004h},
but not the nondimensional deviatoric projection of $\tsr{b}\tsrbar{S}\tsr{b}^2+\tsr{b}^2\tsrbar{S}\tsr{b}\stackrel{\eqref{Eq_TRRSPSR_003c}}{=}\I{b\bar{S}}\tsr{b}^2+\Iab{b}{2}{}{\bar{S}}{}{}\tsr{b}-\III{b}\tsrbar{S}$,
which is \tsc{ch}-reducible by \eqref{Eq_TRRSPSR_003c}. As a consequence, the interpretation $\mathfrak{B}_{(P)}[\tsr{b}, \devnd{S}, \devnd{P}_{\bar\Omega}; \devnd{P}_{\bar S}]$ \eqref{Eq_TRRSPSR_005}
of the classical representation basis in terms of strain-production $P_{\bar S_{ij}}$ \eqref{Eq_TRRSPSR_002d} and rotation-production $P_{\bar\Omega_{ij}}$ \eqref{Eq_TRRSPSR_002e}
is not symmetric in $\devnd{P}_{\bar S}$ \eqrefsab{Eq_TRRSPSR_005d}{Eq_TRRSPSR_002d} and $\devnd{P}_{\bar\Omega}$ \eqrefsab{Eq_TRRSPSR_005e}{Eq_TRRSPSR_002e}.
\begin{table*}[ht]
\begin{center}
\scalebox{0.8}{
\begin{tabular}{lcccccccc}
$\displaystyle\genfrac{}{}{0pt}{}{}{n}\Big\backslash\genfrac{}{}{0pt}{}{m}{}$
          &                         $1$&                     $2$&                     $3$&                   $4$&                 $5$&                $6$&              $7$&            $8$\\\\\hline
       $1$&$1$                         &$0$                     &$0$                     &$0$                   &$0$                 &$0$                &$0$              &$0$            \\
       $2$&$0$                         &$1$                     &$0$                     &$0$                   &$0$                 &$0$                &$0$              &$0$            \\
       $3$&$\dfrac{-\tfrac{1}{18}\dfrac{P_\mathrm{k}}{\rho\varepsilon}-\tfrac{1}{4}\I{b\strd{P}}}{\tfrac{8}{27}+\tfrac{2}{3}\II{b}-\III{b}}$  
                                       &$\dfrac{\tfrac{1}{6}\dfrac{P_\mathrm{k}}{\rho\varepsilon}}{\tfrac{8}{27}+\tfrac{2}{3}\II{b}-\III{b}}$
                                                                &$\dfrac{\tfrac{1}{2}\II{b}-\tfrac{2}{9}}{\tfrac{8}{27}+\tfrac{2}{3}\II{b}-\III{b}}$
                                                                                         &$0$                   &$\dfrac{\tfrac{1}{3}}{\tfrac{8}{27}+\tfrac{2}{3}\II{b}-\III{b}}$                 
                                                                                                                                     &$0$                &$\dfrac{\tfrac{1}{2}}{\tfrac{8}{27}+\tfrac{2}{3}\II{b}-\III{b}}$              
                                                                                                                                                                           &$0$            \\
       ~                                                                                                                                                                                   \\
       $4$&$\dfrac{\tfrac{1}{27}\dfrac{P_\mathrm{k}}{\rho\varepsilon}+\tfrac{1}{6}\I{b\strd{P}}}{\tfrac{8}{27}+\tfrac{2}{3}\II{b}-\III{b}}$  
                                       &$\dfrac{-\tfrac{1}{9}\dfrac{P_\mathrm{k}}{\rho\varepsilon}}{\tfrac{8}{27}+\tfrac{2}{3}\II{b}-\III{b}}$
                                                                &$\dfrac{-\tfrac{2}{3}\II{b}+\tfrac{1}{2}\III{b}}{\tfrac{8}{27}+\tfrac{2}{3}\II{b}-\III{b}}$
                                                                                         &$0$                   &$\dfrac{-\tfrac{2}{9}}{\tfrac{8}{27}+\tfrac{2}{3}\II{b}-\III{b}}$                 
                                                                                                                                     &$0$                &$\dfrac{-\tfrac{1}{3}}{\tfrac{8}{27}+\tfrac{2}{3}\II{b}-\III{b}}$              
                                                                                                                                                                           &$0$            \\
       $5$&$0$                         &$0$                     &$0$                     &$\tfrac{1}{2}$        &$0$                 &$0$                &$0$              &$0$            \\
       $6$&$\dfrac{\big(-\tfrac{1}{9}\II{b}+\tfrac{1}{3}\III{b}-\tfrac{2}{27}\big)\dfrac{P_\mathrm{k}}{\rho\varepsilon}+\big(\tfrac{1}{2}\II{b}+\tfrac{1}{9}\big)\I{b\strd{P}}}{\tfrac{8}{27}+\tfrac{2}{3}\II{b}-\III{b}}$
                                       &$\dfrac{\big(-\tfrac{1}{3}\II{b}-\tfrac{2}{27}\big)\dfrac{P_\mathrm{k}}{\rho\varepsilon}}{\tfrac{8}{27}+\tfrac{2}{3}\II{b}-\III{b}}$
                                                                &$\dfrac{-\II{b}^2+\tfrac{1}{3}\III{b}}{\tfrac{8}{27}+\tfrac{2}{3}\II{b}-\III{b}}$               
                                                                                         &$0$                   &$\dfrac{-\tfrac{1}{3}\II{b}-\tfrac{1}{2}\III{b}}{\tfrac{8}{27}+\tfrac{2}{3}\II{b}-\III{b}}$
                                                                                                                                     &$0$                &$\dfrac{-\II{b}-\tfrac{2}{9}}
                                                                                                                                                                 {\tfrac{8}{27}+\tfrac{2}{3}\II{b}-\III{b}}$
                                                                                                                                                                           &$0$            \\
       $7$&$0$                         &$0$                     &$0$                     &$0$                   &$0$                 &$\tfrac{1}{2}$     &$0$              &$0$            \\
       $8$&$0$                         &$0$                     &$0$                     &$-\tfrac{1}{2}\II{b}$ &$0$                 &$0$                &$0$              &$-\tfrac{1}{2}$\\
\end{tabular}
}\\
\caption{Matrix of coefficients $a_{\tsc{tf}_{nm}}$ for the representation $\devnd{T}_{(n)}=\sum_{m=1}^{8}a_{\tsc{tf}_{nm}}\devnd{F}_{(m)}$ \eqref{Eq_TRRSPSR_010b}
         of the elements $\devnd{T}_{(n)}\in\mathfrak{B}[\tsr{b}, \devnd{S}, \devnd{\Omega}]$ \eqref{Eq_TRRSPSR_004}
         as a linear combination of the elements $\devnd{F}_{(m)}\in\mathfrak{B}[\tsr{b}, \devnd{P}_{\bar S}, \devnd{P}_{\bar\Omega}]$ \eqref{Eq_TRRSPSR_009}.}
\label{Tab_TRRSPSR_005}
\end{center}
\begin{center}
\scalebox{0.65}{
\begin{tabular}{lcccccccc}
$\displaystyle\genfrac{}{}{0pt}{}{}{n}\Big\backslash\genfrac{}{}{0pt}{}{m}{}$
          &                         $1$&                     $2$&                     $3$&                   $4$&                 $5$&                $6$&              $7$&           $8$\\\\\hline
       $1$&$1$                         &$0$                     &$0$                     &$0$                   &$0$                 &$0$                &$0$              &$0$           \\
       $2$&$0$                         &$1$                     &$0$                     &$0$                   &$0$                 &$0$                &$0$              &$0$           \\
       $3$&$\dfrac{-\tfrac{1}{18}\dfrac{P_\mathrm{k}}{\rho\varepsilon}-\tfrac{1}{4}\I{b\strd{P}}}{\tfrac{8}{27}+\tfrac{2}{3}\II{b}-\III{b}}$  
                                       &$\dfrac{\tfrac{1}{6}\dfrac{P_\mathrm{k}}{\rho\varepsilon}}{\tfrac{8}{27}+\tfrac{2}{3}\II{b}-\III{b}}$
                                                                &$\dfrac{\tfrac{1}{4}\II{b}-\tfrac{1}{9}}{\tfrac{8}{27}+\tfrac{2}{3}\II{b}-\III{b}}$
                                                                                         &$\dfrac{\tfrac{1}{4}\II{b}-\tfrac{1}{9}}{\tfrac{8}{27}+\tfrac{2}{3}\II{b}-\III{b}}$
                                                                                                                &$\dfrac{\tfrac{1}{6}}{\tfrac{8}{27}+\tfrac{2}{3}\II{b}-\III{b}}$
                                                                                                                                     &$\dfrac{\tfrac{1}{6}}{\tfrac{8}{27}+\tfrac{2}{3}\II{b}-\III{b}}$
                                                                                                                                                         &$\dfrac{\tfrac{1}{4}}{\tfrac{8}{27}+\tfrac{2}{3}\II{b}-\III{b}}$
                                                                                                                                                                           &$\dfrac{\tfrac{1}{4}}{\tfrac{8}{27}+\tfrac{2}{3}\II{b}-\III{b}}$
                                                                                                                                                                                          \\
       ~                                                                                                                                                                                  \\
       $4$&$\dfrac{\tfrac{1}{27}\dfrac{P_\mathrm{k}}{\rho\varepsilon}+\tfrac{1}{6}\I{b\strd{P}}}{\tfrac{8}{27}+\tfrac{2}{3}\II{b}-\III{b}}$  
                                       &$\dfrac{-\tfrac{1}{9}\dfrac{P_\mathrm{k}}{\rho\varepsilon}}{\tfrac{8}{27}+\tfrac{2}{3}\II{b}-\III{b}}$
                                                                &$\dfrac{-\tfrac{1}{3}\II{b}+\tfrac{1}{4}\III{b}}{\tfrac{8}{27}+\tfrac{2}{3}\II{b}-\III{b}}$
                                                                                         &$\dfrac{-\tfrac{1}{3}\II{b}+\tfrac{1}{4}\III{b}}{\tfrac{8}{27}+\tfrac{2}{3}\II{b}-\III{b}}$
                                                                                                                &$\dfrac{-\tfrac{1}{9}}{\tfrac{8}{27}+\tfrac{2}{3}\II{b}-\III{b}}$
                                                                                                                                     &$\dfrac{-\tfrac{1}{9}}{\tfrac{8}{27}+\tfrac{2}{3}\II{b}-\III{b}}$
                                                                                                                                                         &$\dfrac{-\tfrac{1}{6}}
                                                                                                                                                                 {\tfrac{8}{27}+\tfrac{2}{3}\II{b}-\III{b}}$
                                                                                                                                                                           &$\dfrac{-\tfrac{1}{6}}
                                                                                                                                                                                   {\tfrac{8}{27}+\tfrac{2}{3}\II{b}-\III{b}}$
                                                                                                                                                                                          \\
       $5$&$0$                         &$0$                     &$+\tfrac{1}{4}$         &$-\tfrac{1}{4}$       &$0$                 &$0$                &$0$              &$0$           \\
       $6$&$\dfrac{\big(-\tfrac{1}{9}\II{b}+\tfrac{1}{3}\III{b}-\tfrac{2}{27}\big)\dfrac{P_\mathrm{k}}{\rho\varepsilon}+\big(\tfrac{1}{2}\II{b}+\tfrac{1}{9}\big)\I{b\strd{P}}}{\tfrac{8}{27}+\tfrac{2}{3}\II{b}-\III{b}}$
                                       &$\dfrac{\big(-\tfrac{1}{3}\II{b}-\tfrac{2}{27}\big)\dfrac{P_\mathrm{k}}{\rho\varepsilon}}{\tfrac{8}{27}+\tfrac{2}{3}\II{b}-\III{b}}$
                                                                &$\dfrac{-\tfrac{1}{2}\II{b}^2+\tfrac{1}{6}\III{b}}{\tfrac{8}{27}+\tfrac{2}{3}\II{b}-\III{b}}$               
                                                                                         &$\dfrac{-\tfrac{1}{2}\II{b}^2+\tfrac{1}{6}\III{b}}{\tfrac{8}{27}+\tfrac{2}{3}\II{b}-\III{b}}$               
                                                                                                                &$\dfrac{-\tfrac{1}{6}\II{b}-\tfrac{1}{4}\III{b}}{\tfrac{8}{27}+\tfrac{2}{3}\II{b}-\III{b}}$
                                                                                                                                     &$\dfrac{-\tfrac{1}{6}\II{b}-\tfrac{1}{4}\III{b}}{\tfrac{8}{27}+\tfrac{2}{3}\II{b}-\III{b}}$
                                                                                                                                                         &$\dfrac{-\tfrac{1}{2}\II{b}-\tfrac{1}{9}}
                                                                                                                                                                 {\tfrac{8}{27}+\tfrac{2}{3}\II{b}-\III{b}}$
                                                                                                                                                                           &$\dfrac{-\tfrac{1}{2}\II{b}-\tfrac{1}{9}}
                                                                                                                                                                                   {\tfrac{8}{27}+\tfrac{2}{3}\II{b}-\III{b}}$
                                                                                                                                                                                          \\
       $7$&$0$                         &$0$                     &$0$                     &$0$                   &$\tfrac{1}{4}$      &$-\tfrac{1}{4}$    &$0$              &$0$           \\
       $8$&$0$                         &$0$                     &$-\tfrac{1}{4}\II{b}$   &$+\tfrac{1}{4}\II{b}$ &$0$                 &$0$                &$-\tfrac{1}{4}$  &$\tfrac{1}{4}$\\
\end{tabular}
}\\
\caption{Matrix of coefficients $a_{\tsc{tj}_{nm}}$ for the representation $\devnd{T}_{(n)}=\sum_{m=1}^{8}a_{\tsc{tj}_{nm}}\devnd{J}_{(m)}$ \eqref{Eq_TRRSPSR_010a}
         of the elements $\devnd{T}_{(n)}\in\mathfrak{B}[\tsr{b}, \devnd{S}, \devnd{\Omega}]$ \eqref{Eq_TRRSPSR_004}
         as a linear combination of the elements $\devnd{J}_{(m)}\in\mathfrak{B}[\tsr{b}, \devnd{P}, \devnd{D}]$ \eqref{Eq_TRRSPSR_008}.}
\label{Tab_TRRSPSR_006}
\end{center}
\end{table*}
\begin{figure}[ht!]
\begin{center}
\begin{picture}(240,200)
\put(-10,+10){\includegraphics[angle=0,width=250pt]{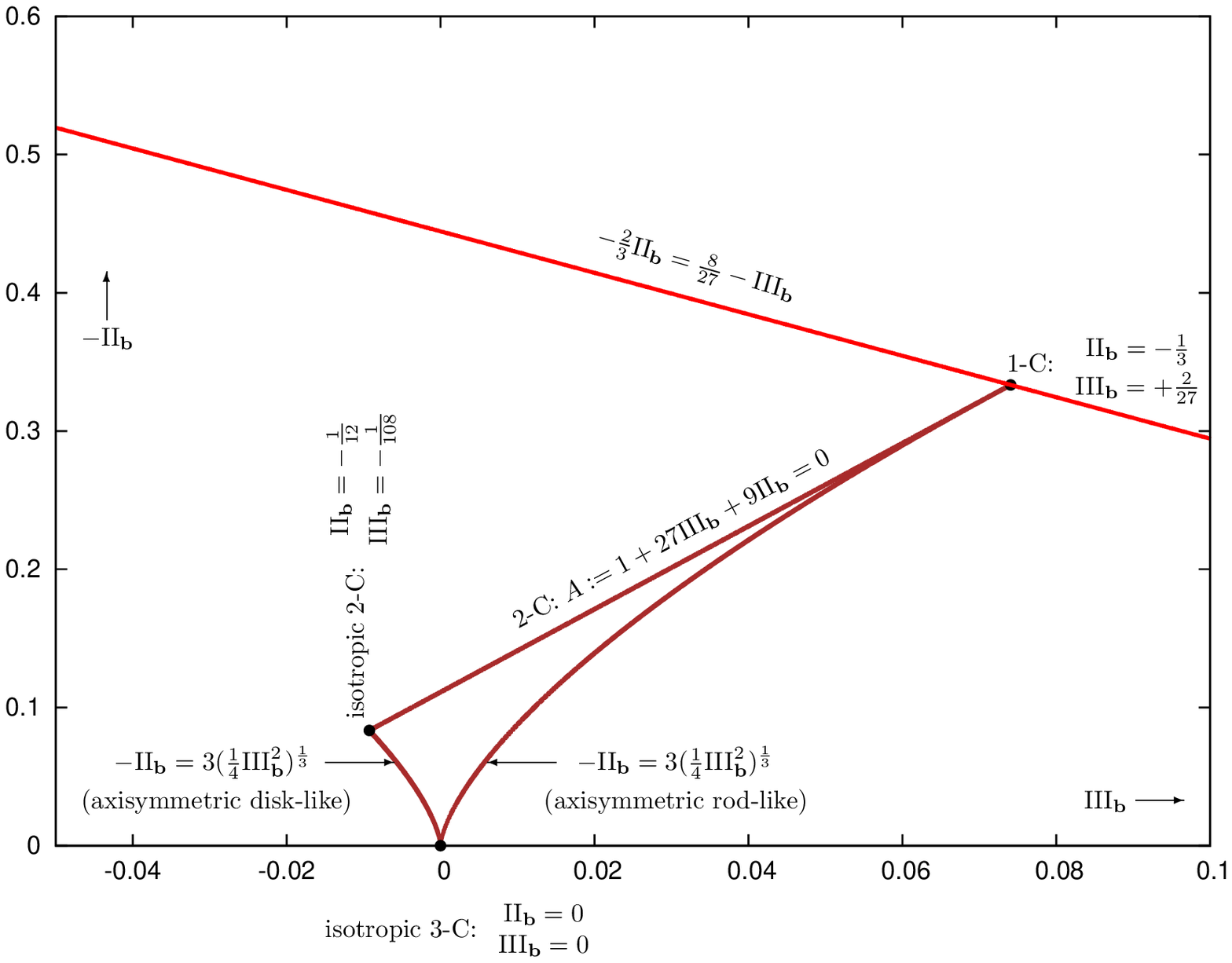}}
\end{picture}
\end{center}
\caption{Locus, in the $(\III{b},-\II{b})$-plane of the invariants of the anisotropy tensor \cite{Lumley_1978a},
         of the condition $-\tfrac{2}{3}\II{b}=\tfrac{8}{27}-\III{b}$ for which the representation $\devnd{T}_{(n)}=\sum_{m=1}^{8}a_{\tsc{tf}_{nm}}\devnd{F}_{(m)}$ \eqref{Eq_TRRSPSR_010b}
         of the mean strain-rate tensor $\devnd{S}$ in $\mathfrak{B}[\tsr{b}, \devnd{P}_{\bar S}, \devnd{P}_{\bar\Omega}]$ would be singular
         (the line $-\tfrac{2}{3}\II{b}=\tfrac{8}{27}-\III{b}$ lies outside of Lumley's realizability triangle \cite{Lumley_1978a,
                                                                                                                     Simonsen_Krogstad_2005a},
         but includes the 1-C point).}
\label{Fig_TRRSPSR_001}
\end{figure}

The basis-elements of $\mathfrak{B}[\tsr{b}, \devnd{P}, \devnd{D}]$ \eqref{Eq_TRRSPSR_008} or $\mathfrak{B}[\tsr{b}, \devnd{P}_{\bar S}, \devnd{P}_{\bar\Omega}]$ \eqref{Eq_TRRSPSR_009}
do not contain explicitly $\bar S_{ij}$ \eqref{Eq_TRRSPSR_002a}. Hence, the practical use of the symmetric bases
$\mathfrak{B}[\tsr{b}, \devnd{P}, \devnd{D}]$ \eqref{Eq_TRRSPSR_008} or $\mathfrak{B}[\tsr{b}, \devnd{P}_{\bar S}, \devnd{P}_{\bar\Omega}]$ \eqref{Eq_TRRSPSR_009},
hinges upon their ability to represent models containing explicitly $\bar S_{ij}$ \eqref{Eq_TRRSPSR_002a} in their formulation, or, in terms of bases,
on the possibility to project the classical representation basis $\mathfrak{B}[\tsr{b}, \devnd{S}, \devnd{\Omega}]$ \eqref{Eq_TRRSPSR_004} onto these alternative bases.
We may readily identify, from \eqrefsabcd{Eq_TRRSPSR_004}{Eq_TRRSPSR_005}{Eq_TRRSPSR_006}{Eq_TRRSPSR_008},
the matrix of coefficients $\underline{\underline{a}}_{(\tsc{jt})}:=[a_{{(\tsc{jt})}_{nm}}]\in\mathbb{R}^{8\times8}$ \tabref{Tab_TRRSPSR_003},
representing the column-vector $\underline{\devnd{J}}:=[\devnd{J}_{(1)},\cdots,\devnd{J}_{(8)}]^\tsc{T}$ \eqref{Eq_TRRSPSR_008}
in terms of the column-vector of the classical basis-elements $\underline{\devnd{T}}:=[\devnd{T}_{(1)},\cdots,\devnd{T}_{(8)}]^\tsc{T}$ \eqref{Eq_TRRSPSR_004},
and from \eqrefsabcd{Eq_TRRSPSR_004}{Eq_TRRSPSR_005}{Eq_TRRSPSR_006}{Eq_TRRSPSR_009},
the matrix of coefficients $\underline{\underline{a}}_{(\tsc{ft})}:=[a_{{(\tsc{ft})}_{nm}}]\in\mathbb{R}^{8\times8}$ \tabref{Tab_TRRSPSR_004},
representing the column-vector $\underline{\devnd{F}}:=[\devnd{F}_{(1)},\cdots,\devnd{F}_{(8)}]^\tsc{T}$ \eqref{Eq_TRRSPSR_005}
in terms of the column-vector of the classical basis-elements $\underline{\devnd{T}}:=[\devnd{T}_{(1)},\cdots,\devnd{T}_{(8)}]^\tsc{T}$ \eqref{Eq_TRRSPSR_004},
\begin{subequations}
                                                                                                                                    \label{Eq_TRRSPSR_010}
\begin{alignat}{6}
\underline{\devnd{J}}=\underline{\underline{a}}_{\tsc{jt}}\;\underline{\devnd{T}} \iff \underline{\devnd{T}}=\underline{\underline{a}}_{\tsc{tj}}\underline{\devnd{J}}
                                                                                                                                    \label{Eq_TRRSPSR_010a}\\
\underline{\devnd{F}}=\underline{\underline{a}}_{\tsc{ft}}\;\underline{\devnd{T}} \iff \underline{\devnd{T}}=\underline{\underline{a}}_{\tsc{tf}}\underline{\devnd{F}}
                                                                                                                                    \label{Eq_TRRSPSR_010b}
\end{alignat}
\end{subequations}
The inverse matrices $\underline{\underline{a}}_{(\tsc{tf})}:=\underline{\underline{a}}_{\tsc{ft}}^{-1}$ \tabref{Tab_TRRSPSR_005}
and $\underline{\underline{a}}_{(\tsc{tj})}:=\underline{\underline{a}}_{\tsc{jt}}^{-1}$ \tabref{Tab_TRRSPSR_006}, which represent the column-vector of the
elements of the classical basis in the 2 symmetric bases \eqref{Eq_TRRSPSR_010}, are readily obtained by straightforward inversion using symbolic calculation \cite{Maxima_20xxa}.
The coefficients \tabrefsab{Tab_TRRSPSR_005}{Tab_TRRSPSR_006} for the representation of the elements of the classical basis $\mathfrak{B}[\tsr{b}, \devnd{S}, \devnd{\Omega}]$ \eqref{Eq_TRRSPSR_004}
in $\mathfrak{B}[\tsr{b}, \devnd{P}, \devnd{D}]$ \eqref{Eq_TRRSPSR_008} and $\mathfrak{B}[\tsr{b}, \devnd{P}_{\bar S}, \devnd{P}_{\bar\Omega}]$ \eqref{Eq_TRRSPSR_009}
are rational (hence continuous except at poles) functions of the invariants. All of the coefficients of $\underline{\underline{a}}_{(\tsc{tf})}$ \tabref{Tab_TRRSPSR_005}
and $\underline{\underline{a}}_{(\tsc{tj})}$ \tabref{Tab_TRRSPSR_006} which could become singular have the same denominator $\tfrac{8}{27}+\tfrac{2}{3}\II{b}-\III{b}$ \tabrefsab{Tab_TRRSPSR_005}{Tab_TRRSPSR_006}.
It turns out that $\tfrac{8}{27}+\tfrac{2}{3}\II{b}-\III{b}\neq0$, because the line $-\II{b}=\tfrac{4}{9}-\tfrac{3}{2}\III{b}$
lies outside of the realizability triangle \cite{Lumley_1978a} of the invariants of $\tsr{b}$, $\II{b}$ and $\III{b}$ \figref{Fig_TRRSPSR_001},
with the unique exception of the 1-C point ($\II{b}=-\tfrac{1}{3}$, $\III{b}=\tfrac{2}{27}$).
Hence, the projection \eqref{Eq_TRRSPSR_010} is valid for any realizable \cite{Lumley_1978a} anisotropy tensor $b_{ij}$, except at the 1-C point.
Using the representation coefficients $a_{(\tsc{tf})_{3m}}$ \tabref{Tab_TRRSPSR_005} and $a_{(\tsc{tj})_{3m}}$ \tabref{Tab_TRRSPSR_006},
the strain-rate tensor is explicitly represented in these bases by
\begin{subequations}
                                                                                                                                    \label{Eq_TRRSPSR_011}
\begin{alignat}{6}
\left(\tfrac{8}{27}+\tfrac{2}{3}\II{b}-\III{b}\right)\dfrac{\mathrm{k} }
                                                           {\varepsilon} \bar S_{ij}
                                                       =& \big(-\tfrac{ 1}
                                                                      {18}\dfrac{P_\mathrm{k}   }
                                                                                {\rho\varepsilon}-\tfrac{1}
                                                                                                        {4}\I{b\strd{P}}\big) b_{ij}
                                                                                                                                    \notag
\end{alignat}
\begin{alignat}{6}
                                                        &+\tfrac{1}
                                                                {6}\dfrac{P_\mathrm{k}   }
                                                                         {\rho\varepsilon}\big(b_{i\ell}b_{\ell j}+\tfrac{2}{3}\II{b}\delta_{ij}\big)
                                                                                                                                    \notag\\
                                                        &+\big(\tfrac{1}{2}\II{b}-\tfrac{2}{9}\big)\dfrac{1}{\rho\varepsilon}\big(P_{\bar S_{ij}}-\tfrac{2}{3}P_\mathrm{k}\delta_{ij}\big)
                                                                                                                                    \notag\\
                                                        &+\tfrac{1}{3}\strd{F}_{(5)ij}+\tfrac{1}{2}\strd{F}_{(7)ij}
                                                                                                                                    \label{Eq_TRRSPSR_011a}\\
\left(\tfrac{8}{27}+\tfrac{2}{3}\II{b}-\III{b}\right)\dfrac{\mathrm{k} }
                                                           {\varepsilon} \bar S_{ij}
                                                       =& \big(-\tfrac{ 1}
                                                                      {18}\dfrac{P_\mathrm{k}   }
                                                                                {\rho\varepsilon}-\tfrac{1}
                                                                                                        {4}\I{b\strd{P}}\big) b_{ij}
                                                                                                                                    \notag\\
                                                        &+\tfrac{1}
                                                                {6}\dfrac{P_\mathrm{k}   }
                                                                         {\rho\varepsilon}\big(b_{i\ell}b_{\ell j}+\tfrac{2}{3}\II{b}\delta_{ij}\big)
                                                                                                                                    \notag\\
                                                        &+\big(\tfrac{1}{4}\II{b}-\tfrac{1}{9}\big)\dfrac{1}{\rho\varepsilon}\big(P_{ij}-\tfrac{2}{3}P_\mathrm{k}\delta_{ij}\big)
                                                                                                                                    \notag\\
                                                        &+\big(\tfrac{1}{4}\II{b}-\tfrac{1}{9}\big)\dfrac{1}{\rho\varepsilon}\big(D_{ij}-\tfrac{2}{3}P_\mathrm{k}\delta_{ij}\big)
                                                                                                                                    \notag\\
                                                        &+\tfrac{1}{6}\strd{J}_{(5)ij}+\tfrac{1}{6}\strd{J}_{(6)ij}
                                                                                                                                    \notag\\
                                                        &+\tfrac{1}{4}\strd{J}_{(7)ij}+\tfrac{1}{4}\strd{J}_{(8)ij}
                                                                                                                                    \label{Eq_TRRSPSR_011b}
\end{alignat}
\end{subequations}
The representation \eqref{Eq_TRRSPSR_011a} of $\bar S_{ij}$ in $\mathfrak{B}[\tsr{b}, \devnd{P}_{\bar S}, \devnd{P}_{\bar\Omega}]$ \eqref{Eq_TRRSPSR_009}
is more compact than the represenation \eqref{Eq_TRRSPSR_011b} in $\mathfrak{B}[\tsr{b}, \devnd{P}, \devnd{D}]$ \eqref{Eq_TRRSPSR_008},
because in the first case the strain-related terms ($\strd{F}_{(3)ij}$, $\strd{F}_{(5)ij}$, $\strd{F}_{(7)ij}$; \eqref{Eq_TRRSPSR_009}) are separated from the
rotation-related terms ($\strd{F}_{(4)ij}$, $\strd{F}_{(6)ij}$, $\strd{F}_{(8)ij}$; \eqref{Eq_TRRSPSR_009}),
whereas in the second case they are coupled because of the identities \eqrefsab{Eq_TRRSPSR_002f}{Eq_TRRSPSR_002g}.

%
%
%
%
%
\section{Change-of-basis relations}\label{TRRSPSR_s_CoBR}
%
%
%
%
%

Since different model proposals in the literature use different bases \cite{Speziale_Sarkar_Gatski_1991a,
                                                                            Ristorcelli_Lumley_Abid_1995a,
                                                                            Fu_Wang_1997a,
                                                                            Naot_Shavit_Wolfshtein_1973a,
                                                                            Launder_Reece_Rodi_1975a,
                                                                            Naot_Shavit_Wolfshtein_1970a,
                                                                            Dafalias_Younis_2009a,
                                                                            Fu_1988a,
                                                                            Craft_Launder_2001a}
it is necessary to establish projection rules for the representation coefficients of various models in different bases.
Any expression (closure) for $\phi_{ij}$ can be represented equivalently in any of the bases,
with linear relations between representation coefficients.
Consider 2 bases with column-vectors of basis-elements $\underline{\devnd{A}}$ and $\underline{\devnd{B}}$, related by
\begin{subequations}
                                                                                                                                    \label{Eq_TRRSPSR_012}
\begin{alignat}{6}
\underline{\devnd{A}}=\underline{\underline{a}}_{(\tsc{ab})}\underline{\devnd{B}}\iff
\underline{\devnd{B}}=\underline{\underline{a}}^{-1}_{(\tsc{ab})}\underline{\devnd{A}}
                                                                                                                                    \label{Eq_TRRSPSR_012a}
\end{alignat}
where the existence of the inverse matric $\underline{\underline{a}}_{(\tsc{ba})}:=\underline{\underline{a}}^{-1}_{(\tsc{ab})}\in\mathbb{R}^{8\times8}$
is equivalent to the linear independence of the basis-elements, and hence to the fact that both $\underline{\devnd{A}}$ and $\underline{\devnd{B}}$ form representation bases.
Denoting $\underline{c}_{\phi\tsc{a}}\in\mathbb{R}^8$ and $\underline{c}_{\phi\tsc{b}}\in\mathbb{R}^8$ the representation coefficients of $\tsr{\phi}$ in each basis,
we readily have
\begin{alignat}{6}
(\rho\varepsilon)^{-1}\tsr{\phi}=&\sum_{n=1}^{8}c_{\phi\tsc{a}_n}\devnd{A}_{(n)}
                                = \underline{c}_{\phi\tsc{a}}^\tsc{t}\underline{\devnd{A}}
                                                                                                                                    \notag\\
\stackrel{\eqref{Eq_TRRSPSR_012a}}{=}&\underline{c}_{\phi\tsc{a}}^\tsc{t}\underline{\underline{a}}_{(\tsc{ab})}\underline{\devnd{B}}
                                   =  \left(\underline{\underline{a}}_{(\tsc{ab})}^\tsc{t}\underline{c}_{\phi\tsc{a}}\right)^\tsc{t}\underline{\devnd{B}}
                                                                                                                                    \label{Eq_TRRSPSR_012b}
\end{alignat}
proving that
\begin{alignat}{6}
\underline{c}_{\phi\tsc{b}}=\underline{\underline{a}}_{(\tsc{ab})}^\tsc{t}\underline{c}_{\phi\tsc{a}}
                                                                                                                                    \label{Eq_TRRSPSR_012c}
\end{alignat}
\end{subequations}
{\em ie} that the representation coefficients in the 2 bases are related by the transpose of the passage-matrix relating the basis-elements.
We may therefore write, using the passage-matrices
$\underline{\underline{a}}_{(\tsc{ht})}\in\mathbb{R}^{8\times8}$ \tabref{Tab_TRRSPSR_001},
$\underline{\underline{a}}_{(\tsc{th})}\in\mathbb{R}^{8\times8}$ \tabref{Tab_TRRSPSR_002},
$\underline{\underline{a}}_{(\tsc{jt})}\in\mathbb{R}^{8\times8}$ \tabref{Tab_TRRSPSR_003},
$\underline{\underline{a}}_{(\tsc{ft})}\in\mathbb{R}^{8\times8}$ \tabref{Tab_TRRSPSR_004},
$\underline{\underline{a}}_{(\tsc{tf})}\in\mathbb{R}^{8\times8}$ \tabref{Tab_TRRSPSR_005}, and
$\underline{\underline{a}}_{(\tsc{tj})}\in\mathbb{R}^{8\times8}$ \tabref{Tab_TRRSPSR_006},
expressing the basis-elements \eqrefsabcd{Eq_TRRSPSR_004}
                                         {Eq_TRRSPSR_005}
                                         {Eq_TRRSPSR_008}
                                         {Eq_TRRSPSR_009}
of any of the bases as a linear combination of the basis-elements of another basis,
\begin{subequations}
                                                                                                                                    \label{Eq_TRRSPSR_013}
\begin{alignat}{6}
\dfrac{\phi_{ij}}
      {\rho\varepsilon}=&\sum_{n=1}^{8}c_{\phi\tsc{t}_n}\strd{T}_{(n)ij}
                       = \sum_{n=1}^{8}c_{\phi\tsc{h}_n}\strd{H}_{(n)ij}
                                                                                                                                    \notag\\
                       =&\sum_{n=1}^{8}c_{\phi\tsc{j}_n}\strd{J}_{(n)ij}
                       = \sum_{n=1}^{8}c_{\phi\tsc{f}_n}\strd{F}_{(n)ij}
                                                                                                                                    \label{Eq_TRRSPSR_013a}
\end{alignat}
\begin{alignat}{6}
\underline{c}_{\phi\tsc{t}}=&\underline{\underline{a}}_{(\tsc{ht})}^\tsc{t}\underline{c}_{\phi\tsc{h}}
                           = \underline{\underline{a}}_{(\tsc{ft})}^\tsc{t}\underline{c}_{\phi\tsc{f}}
                           = \underline{\underline{a}}_{(\tsc{jt})}^\tsc{t}\underline{c}_{\phi\tsc{j}}
                                                                                                                                    \label{Eq_TRRSPSR_013b}\\
\underline{c}_{\phi\tsc{h}}=&\underline{\underline{a}}_{(\tsc{th})}^\tsc{t}\underline{c}_{\phi\tsc{t}}
                                                                                                                                    \label{Eq_TRRSPSR_013c}\\
\underline{c}_{\phi\tsc{j}}=&\underline{\underline{a}}_{(\tsc{tj})}^\tsc{t}\underline{c}_{\phi\tsc{t}}
                                                                                                                                    \label{Eq_TRRSPSR_013d}\\
\underline{c}_{\phi\tsc{f}}=&\underline{\underline{a}}_{(\tsc{tf})}^\tsc{t}\underline{c}_{\phi\tsc{t}}
                                                                                                                                    \label{Eq_TRRSPSR_013e}
\end{alignat}
\end{subequations}

%
%
%
%
%
\section{Isotropic limit}\label{TRRSPSR_s_IL}
%
%
%
%
%

Inspection of the matrices relating different bases \tabrefsatob{Tab_TRRSPSR_001}{Tab_TRRSPSR_006} indicates that the model-coefficient \eqref{Eq_TRRSPSR_013a} of $\bar S_{ij}$ changes depending on the basis used,
and that in the symmetric bases, $\mathfrak{B}[\tsr{b}, \devnd{P}, \devnd{D}]$ \eqref{Eq_TRRSPSR_008} and $\mathfrak{B}[\tsr{b}, \devnd{P}_{\bar S}, \devnd{P}_{\bar\Omega}]$ \eqref{Eq_TRRSPSR_009},
$\bar S_{ij}$ can be represented  as a linear combination of the elements of the basis \eqref{Eq_TRRSPSR_011},
so that
the Rotta-Crow\footnote{\label{ff_TRRSPSR_002}Launder \etal \cite{Launder_Reece_Rodi_1975a} attribute to Crow \cite[(3.6), p. 7]{Crow_1968a} the behaviour of
                        the rapid part of redistribution at the limit of isotropic turbulence,
                        and this has been since repeated by several authors \cite{Fu_Wang_1997a,
                                                                                  Dafalias_Younis_2009a}.
                        Notice however \cite{Naot_Shavit_Wolfshtein_1973a} that the same constraint had also been given by Rotta \cite[p. 558]{Rotta_1951a}.
                       }
constraint \cite{Naot_Shavit_Wolfshtein_1973a,
                 Launder_Reece_Rodi_1975a,
                 Fu_Wang_1997a,
                 Dafalias_Younis_2009a}
can be easily represented by a linear relation between the model coefficients
$\underline{c}_{\phi\tsc{f}}$ \eqrefsab{Eq_TRRSPSR_011a}{Eq_TRRSPSR_013b} or $\underline{c}_{\phi\tsc{j}}$ \eqrefsab{Eq_TRRSPSR_011b}{Eq_TRRSPSR_013b}.
The Rotta-Crow constraint \cite{Naot_Shavit_Wolfshtein_1973a,
                                Launder_Reece_Rodi_1975a,
                                Fu_Wang_1997a,
                                Dafalias_Younis_2009a}
requires that, at the limit of isotropic turbulence, the model for $\phi_{ij}$ should recover the analytical solution $\lim_{\tsr{b}\to0}\phi_{ij}=\tfrac{8}{10}\mathrm{k}\bar S_{ij}=\tfrac{6}{10}\tfrac{4}{3}\mathrm{k}\bar S_{ij}$.
From the representation \eqref{Eq_TRRSPSR_011a} we readily have $\lim_{\tsr{b}\to0}\rho\mathrm{k}\bar S_{ij}=-\tfrac{3}{4}P_{\bar S_{ij}}$. This suggests that an alternative expression of the
Rotta-Crow constraint \cite{Naot_Shavit_Wolfshtein_1973a,
                                Launder_Reece_Rodi_1975a,
                                Fu_Wang_1997a,
                                Dafalias_Younis_2009a}
is
\begin{alignat}{6}
\lim_{\tsr{b}\to0}\phi_{ij}\stackrel{\text{\cite{Rotta_1951a,
                                                 Crow_1968a,
                                                 Naot_Shavit_Wolfshtein_1973a}}}{=}\tfrac{6}{10}\tfrac{4}{3}\mathrm{k}\bar S_{ij}\stackrel{\eqref{Eq_TRRSPSR_011a}}{=}-\tfrac{6}{10}P_{\bar S_{ij}}
                                                                                                                                    \label{Eq_TRRSPSR_014}
\end{alignat}
The practical conclusion from \eqref{Eq_TRRSPSR_014} is that the
Rotta-Crow constraint \cite{Naot_Shavit_Wolfshtein_1973a,
                            Launder_Reece_Rodi_1975a,
                            Fu_Wang_1997a,
                            Dafalias_Younis_2009a}
can be simply included in a closure expressed in $\mathfrak{B}[\tsr{b}, \devnd{P}_{\bar S}, \devnd{P}_{\bar\Omega}]$ \eqref{Eq_TRRSPSR_009}
as a constraint on the model-coefficient of $P^\mathrm{(dev)}_{\bar S_{ij}}$. The implication of this analysis goes beyond the idea of implicit satisfaction \cite{Launder_Reece_Rodi_1975a} of the Rotta-Crow constraint,
as in the isotropization-of-production model \cite{Naot_Shavit_Wolfshtein_1973a}, suggesting that representation bases built using $[\devnd{P}, \devnd{D}]$ or $[\devnd{P}_{\bar S}, \devnd{P}_{\bar\Omega}]$
need not contain explicitly $\bar S_{ij}$.
\begin{figure*}[ht]
\begin{center}
\begin{picture}(500,420)
\put(-5,-20){\includegraphics[angle=0,width=500pt]{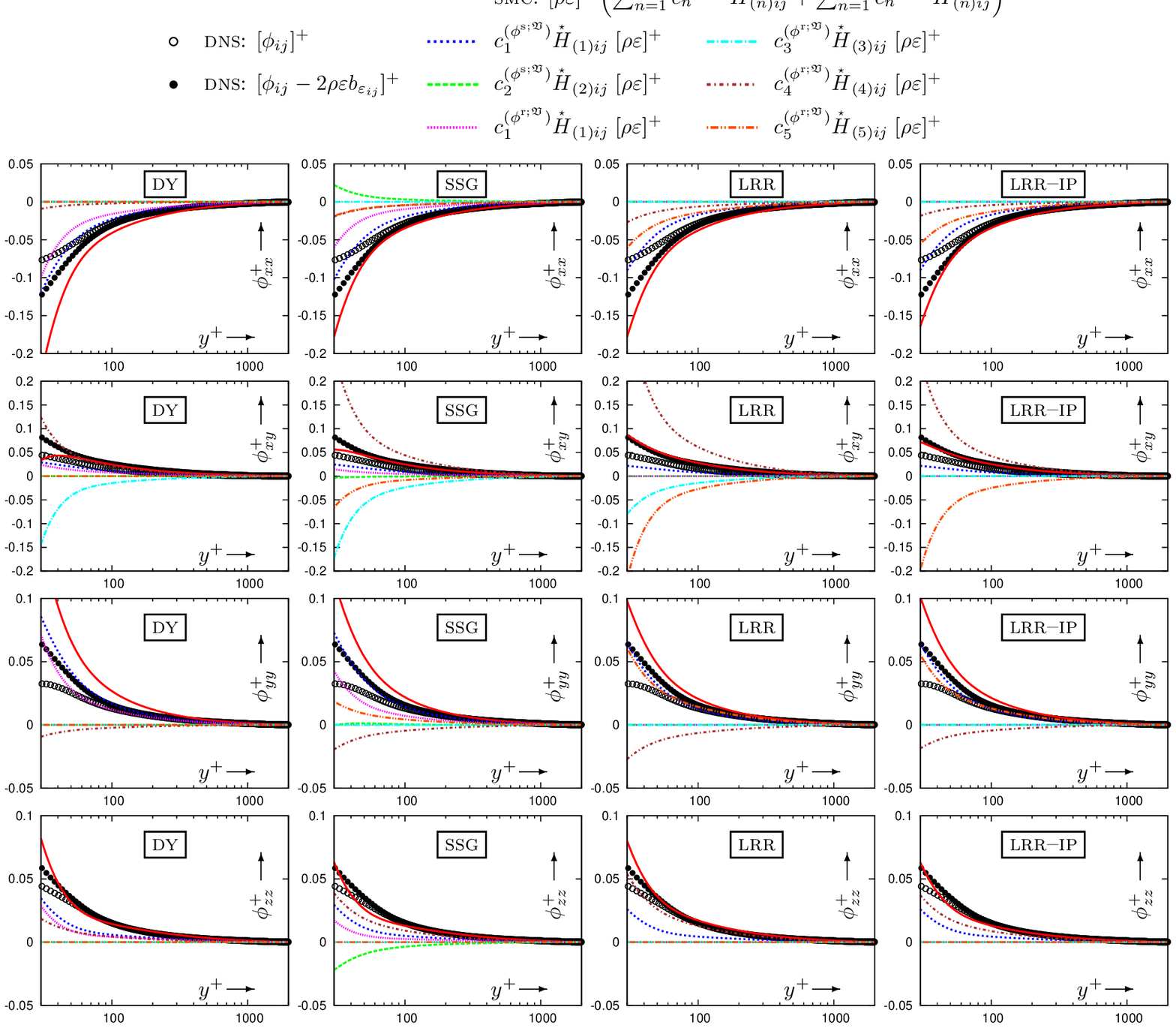}}
\end{picture}
\end{center}
\caption{{\em A priori} term-by-term analysis of \tsc{smc}s (\tsc{lrr} \cite{Launder_Reece_Rodi_1975a},
                                                             \tsc{lrr--ip} \cite{Launder_Reece_Rodi_1975a},
                                                             \tsc{ssg} \cite{Speziale_Sarkar_Gatski_1991a},
                                                             \tsc{dy} \cite{Dafalias_Younis_2009a})
for the components of the redistribution tensor $\phi_{ij}$,
eventually augmented by the anisotropy of dissipation ($\phi_{ij}-2\rho\varepsilon b_{\varepsilon_{ij}}:=\phi_{ij}-\rho(\varepsilon^{(\mu)}_{ij}-\tfrac{2}{3}\varepsilon)$),
with respect to \tsc{dns} data \cite{Hoyas_Jimenez_2008a} for the log and outer regions ($y^+\geq30$) of fully developed incompressible plane channel flow at friction-Reynolds-number $Re_{\tau_w}=2003$
\cite{Hoyas_Jimenez_2008a}.}
\label{Fig_TRRSPSR_002}
\end{figure*}

%
%
%
%
%
\section{Discussion}\label{TRRSPSR_s_D}
%
%
%
%
%

The basic closure for $\phi_{ij}$ (slow return-to-isotropy, \tsc{ri} \cite{Rotta_1951a}, and rapid isotropisation-of-production, \tsc{ip} \cite{Naot_Shavit_Wolfshtein_1970a})
uses the tensors $b_{ij}$ \eqref{Eq_TRRSPSR_002c} and $P_{ij}$ \eqref{Eq_TRRSPSR_002f}, which have a clear physical significance. It was shown in the present work
that the elements of the classic \cite{Ristorcelli_Lumley_Abid_1995a} representation basis $\mathfrak{B}[\tsr{b}, \devnd{S}, \devnd{\Omega}]$ \eqref{Eq_TRRSPSR_004}
can be represented in terms of production by strain $P_{\bar S_{ij}}$ \eqref{Eq_TRRSPSR_002d} or by rotation $P_{\bar\Omega_{ij}}$ \eqref{Eq_TRRSPSR_002e},
as can be interpreted models \cite{Naot_Shavit_Wolfshtein_1973a,
                                   Launder_Reece_Rodi_1975a,
                                   Fu_Wang_1997a}
using their difference $D_{ij}=P_{\bar S_{ij}}-P_{\bar\Omega_{ij}}$ \eqref{Eq_TRRSPSR_002g}.
The widely known \tsc{lrr} \cite{Launder_Reece_Rodi_1975a} and \tsc{ssg} \cite{Speziale_Sarkar_Gatski_1991a} models correspond to different
weightings of the influence of $P_{\bar S_{ij}}$ \eqref{Eq_TRRSPSR_002d} and $P_{\bar\Omega_{ij}}$ \eqref{Eq_TRRSPSR_002e}, the basic \tsc{ip} model \cite{Naot_Shavit_Wolfshtein_1970a}
equally weights both, and the recent \tsc{dy} \cite{Dafalias_Younis_2009a} proposal completely weights out $P_{\bar\Omega_{ij}}$ \eqref{Eq_TRRSPSR_002e}.

To highlight this remark, consider the {\em a priori} term-by-term analysis of different \tsc{smc}s, using \tsc{dns} data \cite{Hoyas_Jimenez_2008a} for fully developed incompressible plane channel flow,
where the contribution of each basis-element to the closure of $\phi_{ij}$ is considered separately \figref{Fig_TRRSPSR_002}.
In this initial comparison, the models were projected in the basis
$\mathfrak{B}_{(P)}[\tsr{b}, \devnd{S}, \devnd{P}_{\bar\Omega}; \devnd{P}_{\bar S}]$ \eqref{Eq_TRRSPSR_005},
which gives the most compact representation, because 3 of the 4 models contain explicitly $\bar S_{ij}$ \cite{Launder_Reece_Rodi_1975a,
                                                                                                              Speziale_Sarkar_Gatski_1991a,
                                                                                                              Dafalias_Younis_2009a}.
Their comparison in the symmetric basis $\mathfrak{B}[\tsr{b}, \devnd{P}_{\bar S}, \devnd{P}_{\bar\Omega}]$ \eqref{Eq_TRRSPSR_009},
will be the subject of future work.
The contribution of $b_{ij}\stackrel{\eqref{Eq_TRRSPSR_005a}}{=:}\strd{H}_{(1)ij}$ to the models was separated \cite{Ristorcelli_Lumley_Abid_1995a}
to a slow part ($\rho\varepsilon\;c^{(\phi^{\mathrm{s};\mathfrak{V}})}_1\strd{H}_{(1)ij}$; \figrefnp{Fig_TRRSPSR_002})
and a rapid part ($\rho\varepsilon\;c^{(\phi^{\mathrm{r};\mathfrak{V}})}_1\strd{H}_{(1)ij}$; \figrefnp{Fig_TRRSPSR_002})
whose coefficient $c^{(\phi^{\mathrm{r};\mathfrak{V}})}_1$ is proportional to $P_\mathrm{k}$ \eqref{Eq_TRRSPSR_006b}, {\em ie} dependent on mean-velocity gradients
(the coefficient of this term is $\neq0$ in the \tsc{ssg} \cite{Speziale_Sarkar_Gatski_1991a} and the \tsc{dy} \cite{Dafalias_Younis_2009a} closures).
For the weakly inhomogeneous (the strongly inhomogeneous near-wall, $y^+\leq30$, region was not plotted; \figrefnp{Fig_TRRSPSR_002}) pure shear flow studied,
the four different \tsc{smc}s (\tsc{lrr} \cite{Launder_Reece_Rodi_1975a},
                               \tsc{lrr--ip} \cite{Launder_Reece_Rodi_1975a},
                               \tsc{ssg} \cite{Speziale_Sarkar_Gatski_1991a},
                               \tsc{dy} \cite{Dafalias_Younis_2009a})
yield similar global results, clearly indicating that they all include the anisotropy of dissipation \figref{Fig_TRRSPSR_002}.
All of the closures use a very similar coefficient for the slow $b_{ij}$ term ($\rho\varepsilon\;c^{(\phi^{\mathrm{s};\mathfrak{V}})}_1\strd{H}_{(1)ij}$; \figrefnp{Fig_TRRSPSR_002}),
which is the basic \tsc{ri} model \cite{Rotta_1951a}. Only the \tsc{ssg} \cite{Speziale_Sarkar_Gatski_1991a} closure contains a nonlinear slow term
($\rho\varepsilon\;c^{(\phi^{\mathrm{s};\mathfrak{V}})}_2\strd{H}_{(2)ij}$), which reduces redistribution by increasing $\phi_{xx}$ (in the present case $\phi_{xx}<0$)
and reducing $\phi_{zz}$ \figref{Fig_TRRSPSR_002}.

On the other hand, the different closures yield similar global predictions for the rapid part of redistribution, but with different weights on each of the 4 basis-elements with rapid coefficients $\neq0$ \figref{Fig_TRRSPSR_002}.
The contributions of $P_{\bar\Omega_{ij}}$ ($\rho\varepsilon\;c^{(\phi^{\mathrm{r};\mathfrak{V}})}_5\strd{H}_{(5)ij}$)
and $P^{(\mathrm{dev})}_{\bar S_{ij}}$ ($\rho\varepsilon\;c^{(\phi^{\mathrm{r};\mathfrak{V}})}_4\strd{H}_{(4)ij}$) to the \tsc{lrr} and \tsc{lrr--ip} closures \cite{Launder_Reece_Rodi_1975a} are quite
similar \figref{Fig_TRRSPSR_002},
the contribution of $\bar S_{xy}\neq0$ ($\rho\varepsilon\;c^{(\phi^{\mathrm{r};\mathfrak{V}})}_3\strd{H}_{(3)ij}$) to the \tsc{lrr} \cite{Launder_Reece_Rodi_1975a} closure for $\phi_{xy}$ along
with the contribution of $P^{(\mathrm{dev})}_{\bar S_{xy}}$ yielding roughly the contribution of $P^{(\mathrm{dev})}_{\bar S_{xy}}$ to the \tsc{lrr--ip} \cite{Launder_Reece_Rodi_1975a} closure,
for which $c^{(\phi^{\mathrm{r};\mathfrak{V}})}_3=0$.
For the \tsc{ssg} \cite{Speziale_Sarkar_Gatski_1991a} closure the contribution of $P_{\bar\Omega_{ij}}$ ($\rho\varepsilon\;c^{(\phi^{\mathrm{r};\mathfrak{V}})}_5\strd{H}_{(5)ij}$)
is roughly $\tfrac{1}{3}$ compared to the \tsc{lrr} and \tsc{lrr--ip} closures \cite{Launder_Reece_Rodi_1975a},
while the contribution of $P^{(\mathrm{dev})}_{\bar S_{ij}}$ ($\rho\varepsilon\;c^{(\phi^{\mathrm{r};\mathfrak{V}})}_4\strd{H}_{(4)ij}$)
is roughly the same \figref{Fig_TRRSPSR_002}. The \tsc{ssg} \cite{Speziale_Sarkar_Gatski_1991a} closure compensates the reduced contribution of $P_{\bar\Omega_{ij}}$ by the rapid
contribution of $b_{ij}$ ($\rho\varepsilon\;c^{(\phi^{\mathrm{r};\mathfrak{V}})}_1\strd{H}_{(1)ij}$), to yield a global prediction for the rapid part of redistribution similar to the
\tsc{lrr} and \tsc{lrr--ip} closures \cite{Launder_Reece_Rodi_1975a} \figref{Fig_TRRSPSR_002}. Nonetheless this is a very different modelling choice,
since the \tsc{ssg} \cite{Speziale_Sarkar_Gatski_1991a} coefficient of the rapid contribution of $b_{ij}$
is proportional to $P_\mathrm{k}\stackrel{\eqref{Eq_TRRSPSR_006b}}{=}\tfrac{1}{2}\Ia{P}{}{\bar S}$, and hence is related to a mechanism
associated with strain-production $P_{\bar S_{ij}}$ and not rotation-production $P_{\bar\Omega_{ij}}$, implying that the \tsc{ssg} closure weights more the former than the latter,
compared to the \tsc{lrr} and \tsc{lrr--ip} closures \cite{Launder_Reece_Rodi_1975a}. Of course this is further amplified in the \tsc{dy} \cite{Dafalias_Younis_2009a} closure \figref{Fig_TRRSPSR_002},
which completely weights out the contribution of $P_{\bar\Omega_{ij}}$ ($\rho\varepsilon\;c^{(\phi^{\mathrm{r};\mathfrak{V}})}_5\strd{H}_{(5)ij}$),
further increasing the rapid contribution of $b_{ij}$ ($\rho\varepsilon\;c^{(\phi^{\mathrm{r};\mathfrak{V}})}_1\strd{H}_{(1)ij}$).

Establishing which is the best model is largely beyond the scope of the present note, requiring systematic comparison for a wide variety of flows.
The purpose of the present comparison is, on the contrary, to highlight the usefulness of element-by-element comparison of different models.

%
%
%
%
%
\section{Extensions}\label{TRRSPSR_s_E}
%
%
%
%
%

To represent $\phi_{ij}$ in strongly inhomogeneous cases, such as wall turbulence, let us assume, as do most models for near-wall turbulence \cite{Gerolymos_Sauret_Vallet_2004a},
that we can identify (either geometrically or through gradients of local turbulence quantities) a unit vector
pointing in the dominant direction of turbulence inhomogeneity, $\vec{e}_\eta$ ($\abs{\vec{e}_\eta}=1$). By adding the deviatoric projection of the tensor product $\vec{e}_\eta\otimes\vec{e}_\eta$,
{\em viz} $\tsr{\eta}:=\vec{e}_\eta\otimes\vec{e}_\eta-\tfrac{1}{3}\tsr{I}_3$, to the 3 basis-generators we obtain (under the condition of linearity in mean-velocity gradients and the fact
that all powers of $\tsr{\eta}$ can be represented as a linear combination of $\tsr{\eta}$ and $\tsr{I}_3$, because by straightforward computation $\tsr{\eta}^2=\tfrac{1}{3}\tsr{\eta}+\tfrac{2}{9}\tsr{I}_3$)
14 more basis-elements \cite{Lo_2011a}, which form, along with the 8 elements of the quasi-homogeneous basis,
a 22-element representation basis for $\phi_{ij}$ in wall turbulence. As an initial application of these ideas, the 22-element basis
obtained by combining $\tsr{\eta}$ with the 3 generators of
$\mathfrak{B}_{(P)}[\tsr{b}, \devnd{S}, \devnd{P}_{\bar\Omega}; \devnd{P}_{\bar S}]:=\big\{\devnd{H}_{(n)},\;n\in\{1,\cdots,8\}\big\}$ \eqref{Eq_TRRSPSR_005}
has been used \cite{Lo_2011a} to represent and compare all known families of single-point \tsc{smc}s in a common basis.

As usual \cite{Speziale_Sarkar_Gatski_1991a,
               Ristorcelli_Lumley_Abid_1995a}
the extension of closures for $\phi_{ij}$ to a reference frame rotating with angular velocity $\vec{\Omega}_\tsc{rf}$ is obtained by replacing everywhere $\bar\Omega_{ij}$ by the intrinsic (absolute) mean rotation-rate
$\bar W_{ij}:=\bar\Omega_{ij}-\epsilon_{ijk}\Omega_{\tsc{rf}_k}$. The extension of incompressible closures to compressible flows is obtained
by replacing the mean strain-rate by its deviatoric projection $\bar S^{(dev)}_{ij}:=\bar S_{ij}-\tfrac{1}{3}\bar S_{\ell\ell}\delta_{ij}$ everywhere,
including in the definition \eqref{Eq_TRRSPSR_001} of $\phi_{ij}$ \cite{Gerolymos_Sauret_Vallet_2004a}.

%
%
%
%
%
\section{Conclusions}\label{TRRSPSR_s_C}
%
%
%
%
%

By splitting the Reynolds-stress production-tensor $P_{ij}$ \eqref{Eq_TRRSPSR_002f}
into strain-production $P_{\bar S_{ij}}$ \eqref{Eq_TRRSPSR_002d} and rotation-production $P_{\bar\Omega_{ij}}$ \eqref{Eq_TRRSPSR_002e}, $P_{ij}\stackrel{\eqref{Eq_TRRSPSR_002f}}{=}P_{\bar S_{ij}}+P_{\bar\Omega_{ij}}$,
the physical interpretation of the tensor $D_{ij}$ \eqref{Eq_TRRSPSR_002g}, present in the original expression of many closures both early \cite{Reynolds_1974a,
                                                                                                                                                 Naot_Shavit_Wolfshtein_1973a,
                                                                                                                                                 Launder_Reece_Rodi_1975a}
and more recent \cite{Fu_Wang_1997a,
                      So_Aksoy_Yuan_Sommer_1996a},
is obtained as the difference between the two mechanisms of production, $D_{ij}\stackrel{\eqref{Eq_TRRSPSR_002g}}{=}P_{\bar S_{ij}}-P_{\bar\Omega_{ij}}$.

The classical representation basis \cite{Ristorcelli_Lumley_Abid_1995a} $\mathfrak{B}[\tsr{b}, \devnd{S}, \devnd{\Omega}]$ \eqref{Eq_TRRSPSR_004}
of closures for pressure-strain redistribution $\phi_{ij}$, can be replaced by equivalent representation bases,
and model coefficients in different bases can be calculated in a systematic way, using passage-matrices (\S\ref{TRRSPSR_s_CoBR}).
Such representation bases, {\em eg} $\mathfrak{B}[\tsr{b}, \devnd{P}_{\bar S}, \devnd{P}_{\bar\Omega}]$ \eqref{Eq_TRRSPSR_009},
are generated exclusively by tensors appearing in the transport equations for the Reynolds-stresses, in which the rate-of-strain tensor $\bar S_{ij}$ \eqref{Eq_TRRSPSR_002a}
is weighted by the anisotropy tensor $b_{ij}$ \eqref{Eq_TRRSPSR_002c}.
In this basis, $\bar S_{ij}$ can be explicitly represented, using the Cayley-Hamilton theorem and its extensions \cite{Rivlin_1955a},
as a linear combination of $b_{ij}$, $P_{\bar S_{ij}}$ and their powers/products which are linear in mean-velicity gradients \eqref{Eq_TRRSPSR_011a},
and the Rotta-Crow constraint \cite{Rotta_1951a,
                                    Crow_1968a}
appears as a limiting constraint on the model-coefficient of $P_{\bar S_{ij}}$.
The practical implication of these results is that we can use the bases $\mathfrak{B}[\tsr{b}, \devnd{P}_{\bar S}, \devnd{P}_{\bar\Omega}]$ \eqref{Eq_TRRSPSR_009} or
$\mathfrak{B}[\tsr{b}, \devnd{P}, \devnd{D}]$ \eqref{Eq_TRRSPSR_008} to compare different models,
including those constructed in the classical basis $\mathfrak{B}[\tsr{b}, \devnd{S}, \devnd{\Omega}]$ \eqref{Eq_TRRSPSR_004},
but also to construct general models for $\phi_{ij}$ which do not contain explicitly the numerically stiff term $\bar S_{ij}$ \cite{Gerolymos_Vallet_2001a}.

An initial term-by-term decomposition on the basis-elements of different models \figref{Fig_TRRSPSR_002}
illustrates how substantially different weighting of various basis-tensors can lead to very similar global results
in weakly inhomogeneous pure shear flow. The detailed element-by-element analysis of different models for various types of flows,
using the bases $\mathfrak{B}[\tsr{b}, \devnd{P}_{\bar S}, \devnd{P}_{\bar\Omega}]$ \eqref{Eq_TRRSPSR_009} or
$\mathfrak{B}[\tsr{b}, \devnd{P}, \devnd{D}]$ \eqref{Eq_TRRSPSR_008},
will be the subject of future work.

\begin{acknowledgment}
The authors are grateful to Profs. M.M. Smith, B.A. Younis and G.F. Smith for enlightening discussions on representation bases.
The authors are listed alphabetically.
\end{acknowledgment}

%
%
%
%
%
%
%
%
%
\bibliographystyle{asmems4}\scriptsize

\begin{thebibliography}{10}

\bibitem{Rotta_1951a}
Rotta, J., 1951.
\newblock ``Statistische {T}heorie nichthomogener {T}urbulenz --- 1.
  {M}itteilung''.
\newblock {\em Z. Phys., {\bf 129}}, pp.~547--572.

\bibitem{Reynolds_1974a}
Reynolds, W.~C., 1974.
\newblock ``Recent advances in the computation of turbulent flows''.
\newblock {\em Adv. Chem. Eng., {\bf 9}}, pp.~193--246.

\bibitem{Lumley_1978a}
Lumley, J.~L., 1978.
\newblock ``Computational modeling of turbulent flows''.
\newblock {\em Adv. Appl. Mech., {\bf 18}}, pp.~123--176.

\bibitem{Speziale_Sarkar_Gatski_1991a}
Speziale, C.~G., Sarkar, S., and Gatski, T.~B., 1991.
\newblock ``Modelling the pressure-strain correlation of turbulence: An
  invariant dynamical systems approach''.
\newblock {\em J. Fluid Mech., {\bf 227}}, pp.~245--272.

\bibitem{Chou_1945a}
Chou, P.~Y., 1945.
\newblock ``On velocity correlations and the solutions of the equations of
  turbulent fluctuations''.
\newblock {\em Quart. Appl. Math., {\bf 3}}, pp.~38--54.

\bibitem{Ristorcelli_Lumley_Abid_1995a}
Ristorcelli, J.~R., Lumley, J.~L., and Abid, R., 1995.
\newblock ``A rapid-pressure covariance representation consistent with the
  {T}aylor-{P}roudman theorem materially frame indifferent in the {2-D}
  limit''.
\newblock {\em J. Fluid Mech., {\bf 292}}, pp.~111--152.

\bibitem{Fu_Wang_1997a}
Fu, S., and Wang, C., 1997.
\newblock ``Second-moment closure modelling of turbulence in a noninertial
  frame''.
\newblock {\em Fluid Dyn. Res., {\bf 20}}, pp.~43--65.

\bibitem{Naot_Shavit_Wolfshtein_1973a}
Naot, D., Shavit, A., and Wolfshtein, M., 1973.
\newblock ``2-point-correlation model and the redistribution of
  {R}eynolds-stresses''.
\newblock {\em Phys. Fluids, {\bf 16}}(6), June, pp.~738--743.

\bibitem{Launder_Reece_Rodi_1975a}
Launder, B.~E., Reece, G.~J., and Rodi, W., 1975.
\newblock ``Progress in the development of a {R}eynolds-stress turbulence
  closure''.
\newblock {\em J. Fluid Mech., {\bf 68}}, pp.~537--566.

\bibitem{Launder_1989a}
Launder, B.~E., 1989.
\newblock ``Second-moment closure: Present...and future?''.
\newblock {\em Int. J. Heat Fluid Flow, {\bf 10}}, pp.~282--300.

\bibitem{Naot_Shavit_Wolfshtein_1970a}
Naot, D., Shavit, A., and Wolfshtein, M., 1970.
\newblock ``Interactions between components of the turbulent velocity
  correlation tensor due to pressure fluctuations''.
\newblock {\em Israel J. Techn., {\bf 8}}(3), pp.~259--269.

\bibitem{Dafalias_Younis_2009a}
Dafalias, Y.~F., and Younis, B.~A., 2009.
\newblock ``Objective model for the fluctuating pressure-strain-rate
  correlations''.
\newblock {\em ASCE J. Eng. Mech., {\bf 135}}(9), Sept., pp.~1006--1014.

\bibitem{Fu_1988a}
Fu, S., 1988.
\newblock ``Computational modelling of turbulent swirling flows with
  second-moment closures''.
\newblock Phd thesis, University of Manchester Institute of Science and
  Technology, Manchester [{\sc gbr}].

\bibitem{Craft_Launder_2001a}
Craft, T.~J., and Launder, B., 2001.
\newblock ``Principles and performance of {TCL}-based second-moment closures''.
\newblock {\em Flow Turb. Comb., {\bf 66}}, pp.~355--372.

\bibitem{Lo_2011a}
Lo, C., 2011.
\newblock ``Fermeture de la turbulence au second-ordre proche paroi bas\'ee sur
  l'analyse des donn\'ees {\sc dns}''.
\newblock Doctorat, Universit\'e Pierre-et-Marie-Curie, Paris {[{\sc fra}]},
  Sept.

\bibitem{Shima_1998a}
Shima, N., 1998.
\newblock ``Low-{R}eynolds-number second-moment closure without wall-reflection
  redistribution terms''.
\newblock {\em Int. J. Heat Fluid Flow, {\bf 19}}, pp.~549--555.

\bibitem{Rivlin_1955a}
Rivlin, R.~S., 1955.
\newblock ``The theory of matrix polynomials and its application to the
  mechanics of isotropic continua''.
\newblock {\em Indiana Univ. Math. J., {\bf 4}}, pp.~681--702.

\bibitem{Spencer_Rivlin_1959a}
Spencer, A. J.~M., and Rivlin, R.~S., 1959.
\newblock ``The theory of matrix polynomials and its application to the
  mechanics of isotropic continua''.
\newblock {\em Arch. Rat. Mech. Anal., {\bf 2}}, pp.~309--336.

\bibitem{Smith_1971a}
Smith, G.~F., 1971.
\newblock ``On isotropic functions of symmetric tensors, skew-symmetric tensors
  and vectors''.
\newblock {\em Int. J. Eng. Sci., {\bf 9}}, pp.~899--916.

\bibitem{Rivlin_Ericksen_1955a}
Rivlin, R.~S., and Ericksen, J.~L., 1955.
\newblock ``Stress-deformation relations for isotropic materials''.
\newblock {\em Indiana Univ. Math. J., {\bf 4}}, pp.~323--425.

\bibitem{Gerolymos_Sauret_Vallet_2004a}
Gerolymos, G.~A., Sauret, E., and Vallet, I., 2004.
\newblock ``Contribution to the single-point-closure {R}eynolds-stress
  modelling of inhomogeneous flow''.
\newblock {\em Theor. Comp. Fluid Dyn., {\bf 17}}(5--6), Sept., pp.~407--431.

\bibitem{Gibson_Launder_1978a}
Gibson, M.~M., and Launder, B.~E., 1978.
\newblock ``Ground effects on pressure fluctuations in the atmospheric
  boundary-layer''.
\newblock {\em J. Fluid Mech., {\bf 86}}, pp.~491--511.

\bibitem{So_Aksoy_Yuan_Sommer_1996a}
So, R. M.~C., Aksoy, H., Yuan, S.~P., and Sommer, T.~P., 1996.
\newblock ``Modeling {R}eynolds-number effects in wall-bounded turbulent
  flows''.
\newblock {\em ASME J. Fluids Eng., {\bf 118}}, June, pp.~260--267.

\bibitem{Simonsen_Krogstad_2005a}
Simonsen, A.~J., and Krogstad, P.~{\AA}., 2005.
\newblock ``Turbulent stress invariant analysis: Classification of existing
  terminology''.
\newblock {\em Phys. Fluids, {\bf 17}}, pp.~088103(1--4).

\bibitem{Maxima_20xxa}
{\sc maxima}.
\newblock A computer algebra system.
\newblock {\color{blue}{http://maxima.sourceforge.net}}.

\bibitem{Crow_1968a}
Crow, S.~C., 1968.
\newblock ``A suggestion for the numerical computation of the steady
  {N}avier-{S}tokes equations''.
\newblock {\em J. Fluid Mech., {\bf 33}}, pp.~1--20.

\bibitem{Hoyas_Jimenez_2008a}
Hoyas, S., and Jim\'enez, J., 2008.
\newblock ``{R}eynolds number effects on the {R}eynolds-stress budgets in
  turbulent channels''.
\newblock {\em Phys. Fluids, {\bf 20}}, pp.~101511(1--8).

\bibitem{Gerolymos_Vallet_2001a}
Gerolymos, G.~A., and Vallet, I., 2001.
\newblock ``Wall-normal-free near-wall {R}eynolds-stress closure for {3-D}
  compressible separated flows''.
\newblock {\em AIAA J., {\bf 39}}(10), Oct., pp.~1833--1842.

\end{thebibliography}

\normalsize
%
%
%
%
%
%
%
%
%

\end{document}